\documentstyle[prb,aps,multicol,amsmath,graphicx,psfrag]{revtex}

\begin{document}
\draft

\newcommand{\fig}[2]{\includegraphics[width=#1]{Figures/#2}}
\newcommand{\diag}[3]{
        \raisebox{-#1}{
        \includegraphics[width=#2]{Figures/#3.eps}}}
 \newcommand{\startlargeeq}{
         \end{multicols}\widetext%
                \vspace{-.5cm}\noindent\rule{8.8cm}{.1mm}%
                                       \rule{.1mm}{.4cm}}
 \newcommand{\stoplargeeq}{
         \vspace{-.5cm}\noindent\rule{9.1cm}{0mm}%
                                \rule{.1mm}{.4cm}%
                                \rule[.4cm]{8.8cm}{.1mm}%
         \begin{multicols}{2} }

\title{Field theory for generalized Shastry-Sutherland models}
\author{David Carpentier \\ Institute for Theoretical Physics,
  University of California, Santa Barbara, CA 93106\\
and \\
CNRS-Laboratoire de Physique de l'ENS-Lyon, 46 All{\'e}e d'Italie, 69007
Lyon, France}
\author{Leon Balents \\  Physics Department, University of
  California, Santa Barbara, CA 93106}

\date{\today}

\maketitle

\begin{abstract}
  We consider the bosonic  dimer representation for {\it Generalized
    Shastry-Sutherland models} that have the same symmetries as the
  original Shastry-Sutherland model and preserve the exact dimer
  eigenstate.  Various phases with differing types of magnetic order
  are found within mean-field theory for the corresponding low-energy
  effective dimer field theory. Transitions are allowed between any of
  these mean-field phases, which are {\sl dimer bose condensates}, and
  with the dimer phase, which is the {\sl dimer bose vacuum}.  The
  N{\'e}el state, absent from this mean-field study, is described as a
  bosonic Mott insulator induced by the coupling to the underlying
  lattice. Moreover, dimer bose condensates with {\sl local} N{\'e}el
  order are found to be unstable to spiral states.  Instead of a
  direct phase transition between the dimer and the N{\'e}el phases,
  we propose an intermediate weakly incommensurate spin-density wave
  (WISDW) phase.  The stability of the mean-field transitions is
  studied by renormalization techniques in $d=2$, the upper
  critical dimension.  While the transition from the N{\'e}el phase is
  found to be stable, the transition point from the dimer phase is not
  perturbatively accessible.  We argue that the latter renormalization
  results point to the possibility of an intermediate phase of a
  different kind.   
\end{abstract}

\vspace{0.15cm}

\begin{multicols}{2}
\section{Introduction}
\label{sec:intro}

The Shastry-Sutherland (SS) model\cite{Shastry81} is a remarkable
two-dimensional (2D)  analog of 
the Majumdar-Gosh spin chain\cite{Majumdar69}, possessing an exact
dimerized 
eigenstate, despite non-trivial spin-spin interactions. It recently
received much attention\cite{Miyahara98,Singh99,Koga00,Miyahara00}
 due to its relevance for the description of
the quasi-two dimensional compound\cite{exp} SrCu$_{2}$(BO$_{3}$)$_{2}$. 
 The original model is described by the Hamiltonian\cite{notations}
\begin{equation}\label{SS-H}
  H_{SS}=J_{1}\sum_{\langle i,j\rangle} \vec{S}_{i}\cdot\vec{S}_{j}
  +J_{2}\sum_{\langle\langle k,l\rangle\rangle}
  \vec{S}_{k}\cdot\vec{S}_{l}, 
\end{equation}
where $J_{1}$ is the coupling along the solid lines of figure
\ref{fig:ss1}, and  
$J_{2}$ along the dashed bonds.  
 As both interaction terms couple the two spins
of each dimer (solid bond in Fig.~\ref{fig:ss1}) {\sl symmetrically},
the dimerized state (direct-product of singlet states on each dimer) 
is an exact eigenstate of the hamiltonian
(\ref{SS-H}) for any value of $J_{1}$ and $J_{2}$. 
 
In this paper, we will also consider the three-dimensional (3D)
extension of the SS model\cite{Miyahara00}, relevant to the
description of the SrCu$_{2}$(BO$_{3}$)$_{2}$ compound. This 3D model
is defined on a 3D lattice consisting of a stack of alternate 2D SS
lattices (see Fig.\ref{fig:ss1} : every other layer in the $z$
direction is rotated by $\pi/2$).  The spins of dimers lying on top of each
other interact via an additional coupling,
\begin{equation}
  H_{3d} = J_{3}~ \sum_{z}\sum_{(i,j)} \vec{S}_{i,z}\cdot\vec{S}_{j,z+1},
\end{equation}  
where $i,j$ span the spins of the two dimers on top of each other
(see Fig. \ref{fig:ss1}). The dimerized SS is also an exact eigenstate
of this generalization.

\begin{figure}[htb]
\centerline{
\psfrag{a}{a}\psfrag{b}{b}\psfrag{c}{c}\psfrag{d}{d}
\psfrag{a1}{${\bf a}_{1}$}\psfrag{a2}{${\bf a}_{2}$}
\fig{4cm}{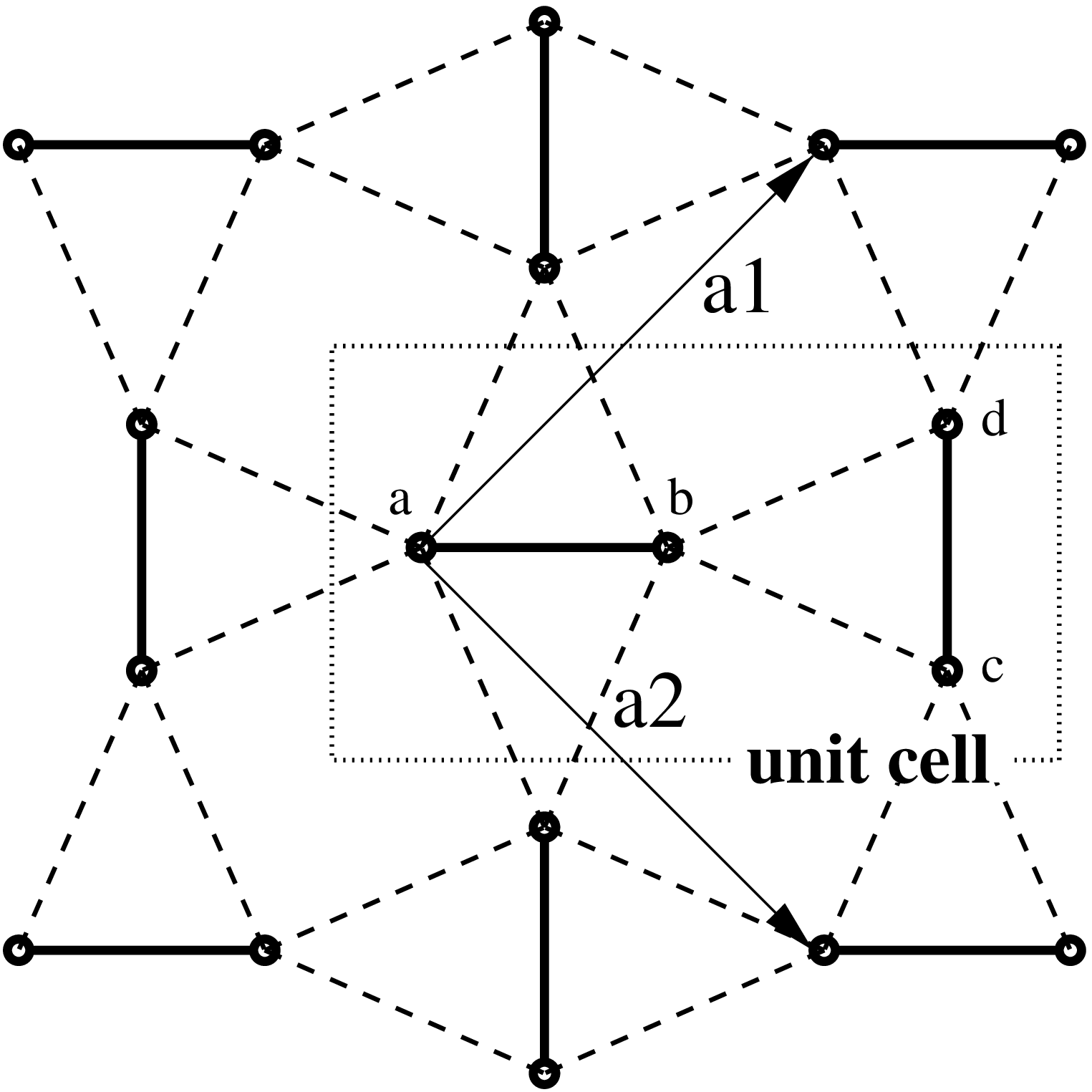}
\hspace{.5 cm}
\fig{3cm}{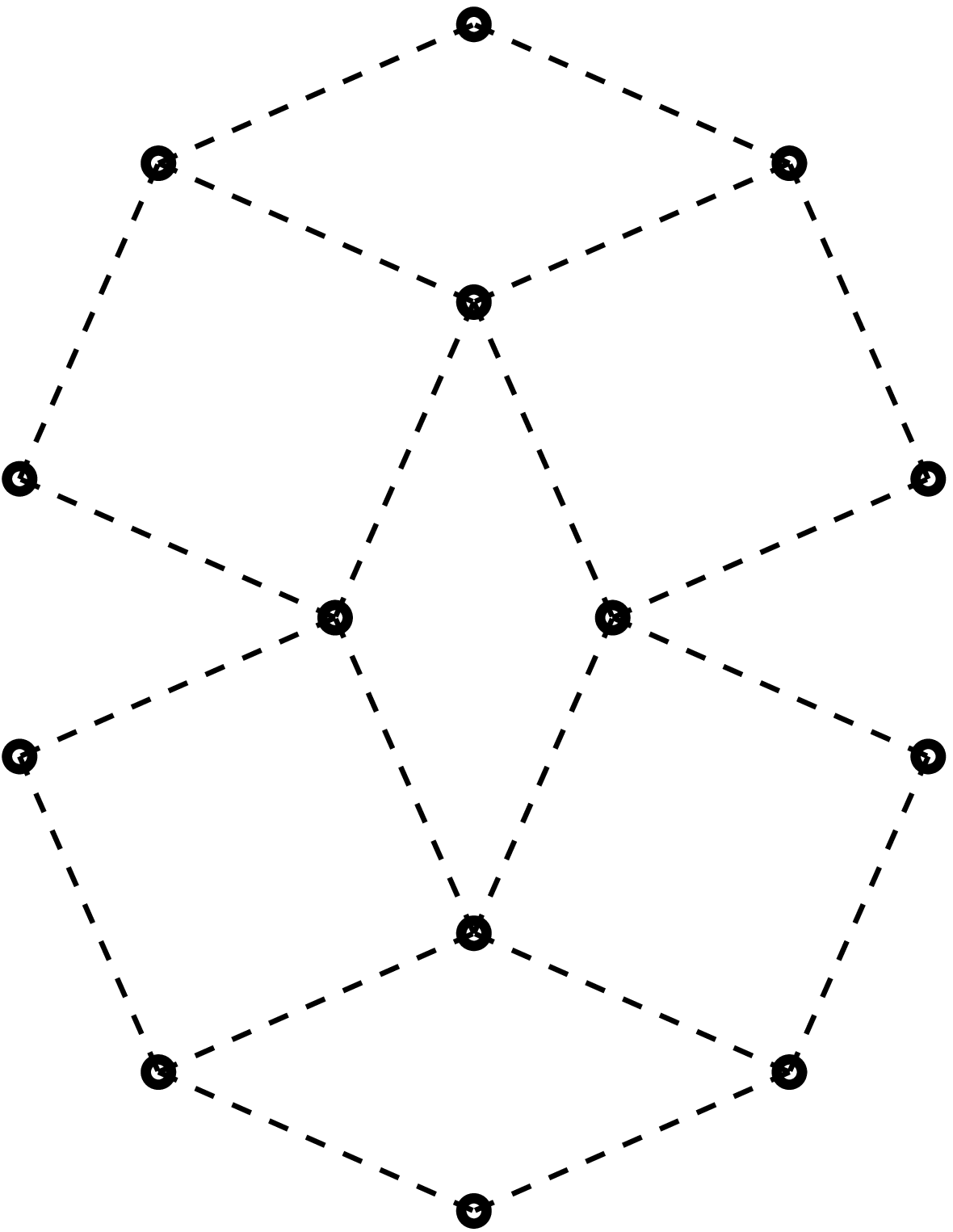}
}
\centerline{
\psfrag{J3}{$J_{3}$}
\psfrag{upper layer}{upper layer}\psfrag{lower layer}{lower layer}
\fig{6cm}{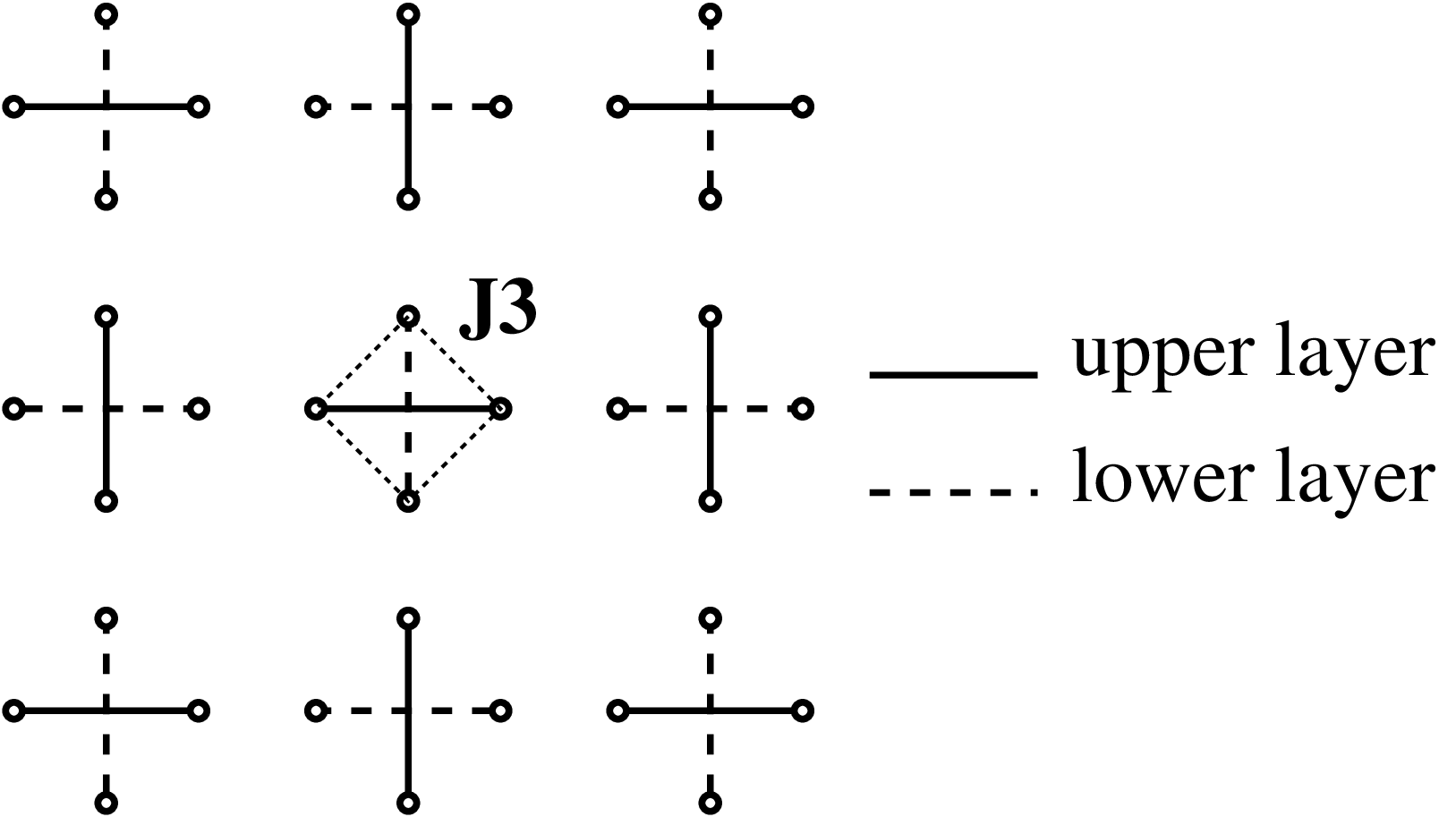}
}
\vspace{.3cm}
\caption{Topology of the lattice defining the Shastry-Sutherland model
(top left). The right lattice, equivalent to a square lattice, corresponds
to the situation where $J_{1}=0$.
 The bottom figure shows the spatial configuration of the layers in
the 3D version of the Shastry-Sutherland model. 
}
\label{fig:ss1}
\end{figure}    

The ground state of the SS model is known in the two extreme limits.
For $J_{2}=0$, the dimers are completely decoupled from each other and
the above SS dimer state is obviously the ground state. As all other
states have an excitation energy of order $J_1$, we expect the SS
dimer state to remain the ground state for small $J_{2}\ll J_{1}$.  On
the other hand for $J_1=0$, the SS model becomes equivalent to the
Heisenberg model with the exchange coupling $J_2$ defined on a square
lattice as can be seen on Fig.\ref{fig:ss1} (right upper part). In
this limit the ground state is a N{\'e}el state, possessing staggered
magnetization\cite{comment-Heinsenberg} (on the underlying square
lattice). 
  The additional $J_1$
interaction in this picture corresponds to a staggered diagonal
exchange introduced in half the plaquettes.  This dimerization {\sl
  quadruples} the unit cell, which thus contains four spins.  A
convenient choice of unit cell is shown in Fig.\ref{fig:ss1}.  For
$J_1 \ll J_2$, the N{\'e}el state is expected to persist, as can be
verified in perturbation theory\cite{Singh99}.  
The SS model is thus expected to have 
 an ``antiferromagnetic'' (in the square lattice sense) 
ground state for $J_2/J_1> {\cal J}_c^\prime$ and 
the exact dimerized ground state for $J_2/J_1
< {\cal J}_c \leq {\cal J}_c^\prime$.  The intervening range has been
investigated by a number of
authors\cite{Albrecht95,Miyahara98,Singh99} .  Various numerical
methods\cite{Singh99,Miyahara98}\ suggest that ${\cal J}_c \approx
{\cal J}_c^\prime \approx 0.69$, and possibly a direct transition
between the two states.  Series expansions about the dimer limit show
a vanishing of the triplet excitation gap very close to this point,
and a second-order transition of some type was suggested. Possible
intermediate spiral phases were discussed within a mean-field slave
boson approach\cite{Albrecht95}. 

The purpose of the present paper is to propose a new framework to
investigate the transitions and possible phases of {\sl generalized
  Shastry-Sutherland} (GSS) models in 2D and 3D.  By {\sl generalized
  Shastry-Sutherland} models we mean spin Hamiltonians defined on the
same lattice as the original SS model (Fig.\ref{fig:ss1}), but with
general local interactions that satisfy two conditions: (1) the SS
dimer state remains an exact eigenstate of the model, and (2) the
symmetries of the SS model are maintained.  To study the phases of
these models, we take the SS dimer state as a starting point, and
construct a field theory perturbatively around it. The underlying idea
of our approach is the following : as the SS dimer state is an exact
non-critical eigenstate, if the system was undergoing a second-order
quantum phase transition from it, the ground-state would be
non-critical on one side of the transition. The similarity between
this scenario and the two-dimensional superfluid-insulator
transition\cite{Fisher89} motivates the introduction of a {\sl bosonic
  dimer spin representation} which is done in section
\ref{sec:models}.  This method represents the four states of a pair of
spins by the vacuum and three singly-occupied states of a triplet
  of hard-core ``dimer-bosons''.  The GSS spin model is then exactly translated
into a square lattice boson model with complex interactions, whose
boson vacuum corresponds to the SS dimer state.

Focusing on the universal
properties of the possible transitions of this bosonic model, we
derive the corresponding low-energy, long wavelength effective dimer
field theory 
(DFT).  In constructing this field theory, we
assume that all ordering occurs for small crystal momentum $k$ on the
scale of the Brillouin zone, i.e. $ka \ll 1$, where $a$ is the lattice
spacing.  Because of the four-site unit cell, this includes {\sl both}
the N{\'e}el and dimer states, neither of which breaks the translational
symmetry of the SS lattice.  The upper critical dimension of this new DFT
is found to be $d=2$.  Hence, in a sufficiently three dimensional
material, the phases of the GSS model can be described within a
Mean-Field Theory (MFT).  The phases captured within this mean-field
approach, which correspond to various coherent {\sl condensates} of
the dimer bosons, are described in section \ref{sec:smmft}. Due to the
intrinsic complexity of the DFT, the mean-field description is carried
on a simpler 2D ``sub-model'' whose analysis contains the necessary phases.
Besides the dimer state, several phases with interesting local
magnetic order are found within this approach : (i) an antiparallel phase
where on each dimer the spins arrange in an antiparallel way, (ii) a chiral
phase where these spins rotate with respect to each other within a
bond but the average magnetization vanishes, and (iii) a spiral phase with
both non-zero magnetization and chirality. Mean field theory predicts
continous transitions between all these phases except between the
chiral and the spiral phases, connected by a first-order transition.
 
Surprisingly, {\sl none} of these bose condensates corresponds to the
antiferromagnetic (on the associated square lattice) phase!  On the
contrary, we show that (for the full set of GSS models), if the
quantum phase transition out of the dimerized state can be regarded as
Bose condensation (even if it is {\sl first order}), the resulting
ordered state {\sl cannot} have N{\'e}el order without fine-tuning of
parameters.  The closest such a Bose condensate can approach the
antiferromagnet is to sustain weakly {\sl incommensurate} spin-density
wave (WISDW) order (i.e. a periodic modulation of the expectation
value of the total spin on each dimer) at a wavevector near but not
equal to $(\pi,\pi)$ on the underlying square lattice.  Moreover, even
such a WISDW state necessarily contains concommitent transverse (to
the local WISDW quantization axis) magnetic (e.g. antiparallel) and
chiral order.  These two properties follow from symmetry
considerations.  In particular, the total spin on each dimer is
invariant under spatial reflections, and hence must be bilinear in the
Bose fields.  Moreover, since it is an SU(2) vector, it can only be
related to a cross-product of these fields, which must themselves
therefore have non-zero values along the two transverse axes.  The
latter condition implies spontaneously broken reflection and
(four-fold) rotational symmetry, and full breaking of SU(2) symmetry
(i.e. with no remaining invariant subgroups).  These conditions lead
to a generic long-wavelength instability of the gapless magnon
(Goldstone) modes of the broken SU(2) to incommensurate ordering.

The upshot is that to describe the ordinary antiferromagnet, with
commensurate N{\'e}el order and no transverse magnetic ordering, it is
necessary to go beyond the simplest DFT and include the couplings
between the dimer bosons and the underlying lattice.  We show that, in
this context, the antiferromagnetic phase is described not as a Bose
condensate but as a bosonic Mott insulator.  The failure of mean-field
theory to describe such a Mott insulator can be understood from the
uncertainty principle.  Because the Mott insulator is a state with
``definite'' boson number, the strong fluctuations of the conjugate
boson phase invalidate the mean-field approach.  Simple arguments then
show that a direct continuous transition from this antiferromagnetic
phase to the dimer SS state is not possible.

\begin{figure}[htb]
\psfrag{a}{$a$}\psfrag{b}{$b$}\psfrag{a'}{$a'$}\psfrag{b'}{$b'$}
\centerline{
\fig{4cm}{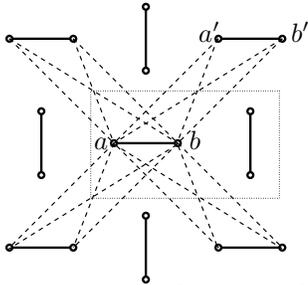}}
\caption{Four-spin interaction that is {\sl exactly} equivalent to a
  dimer-boson hopping term.  The dashed lines illustrate interactions
  between the $a$ and $b$ spins of neighboring unit cells (see
  Eq.~\ref{eq:newints}).  There are identical interactions, rotated by
  forty-five degrees, between the $c$ and $d$ spins (not shown). }
\label{fig:newints}
\end{figure}     

Instead, in three dimensions the theory suggests it is possible to
cross from the Dimer to N{\'e}el states via an {\sl intermediate} bose
condensate in a sequence of continuous transitions.  At the first
transition the dimer Bosons condense to form a WISDW state.  Upon
further varying some parameter in the theory, the mean density of
dimer bosons approaches one per dimer, and the system undergoes a {\sl
  superfluid-insulator}-like transition to the bosonic Mott insulator
which corresponds to the N{\'e}el state.  While there is no numerical
evidence for this two-stage transition in the usual SS model, it
should occur naturally in other models in the GSS class.  The
dimer-boson representation gives strong hints as to the nature of the
additional interactions that should be added to the SS model in order
to facilitate such a two-step transition.  In particular, by a
judicious choice of {\sl four-spin} interactions, the {\sl kinetic
  energy} of the dimer-bosons can be explicitly increased.  This is
accomplished by the following term, illustrated in
Fig.~\ref{fig:newints}:
\begin{eqnarray}
  H^* & = & -J^* \sum_{\langle {\bf xx'}\rangle} \bigg[ 
      \frac{1}{2} (\vec{S}_a-\vec{S}_b)\cdot(\vec{S}_{a'} -
      \vec{S}_{b'}) \nonumber \\
      & & + 2 ( \vec{S}_a\times
      \vec{S}_b)\cdot(\vec{S}_{a'}\times\vec{S}_{b'}) + \left(aba'b'
        \leftrightarrow cdc'd'\right)
  \bigg],  \label{eq:newints}
\end{eqnarray}
where the subscripts, $abcd,a'b'c'd'$, indicate the appropriate spin
in the unit cell centered on {\bf x,x'}, respectively.  The sum is
taken over nearest-neighbor unit cells, i.e. all pairs of lattice
vectors satisfying ${\bf x-x'} = \pm {\bf a}_1, \pm {\bf a}_2$, where
${\bf a}_{1,2}$ are the primitive vectors shown in Fig.~\ref{fig:ss1}.
Remarkably, this rather strange interaction can be exactly rewritten,
using the formulae in section II, as a simple nearest-neighbor
dimer-boson hopping term.  It therefore preserves the exact dimer
eigenstate.  In the original SS model $J^*=0$, and the boson kinetic
energy arises only through high-order virtual processes involving many
$J_2$ exchanges.  By explicitly including it, $H_{SS} \rightarrow
H_{GSS} = H_{SS}+H^*$, the bosons acquire a ``bare'' kinetic energy
and Bose condensate is rendered more favorable -- indeed if $J^*$ is
increased and a low density of bosons maintained by simultaneously
increasing $J_1$, the transition out of the dimer state becomes
parametrically better described as condensation of a dilute,
weakly-interacting Bose gas.  A possible schematic phase diagram
showing the evolution of the ground state of the GSS model on
increasing $J^*$ in three dimensions is indicated in
Fig.~\ref{fig:schempds}a.

\begin{figure}[htb]
\psfrag{(a)}{\bf (a)}
\psfrag{WISDW}{\bf WISDW}
\psfrag{Neel}{\bf N{\'e}el}
\psfrag{dimer} {\bf dimer}
\psfrag{J2/J1}{$J_2/J_1$}
\psfrag{J*/J2}{$J^*/J_2$}
\centerline{\fig{6cm}{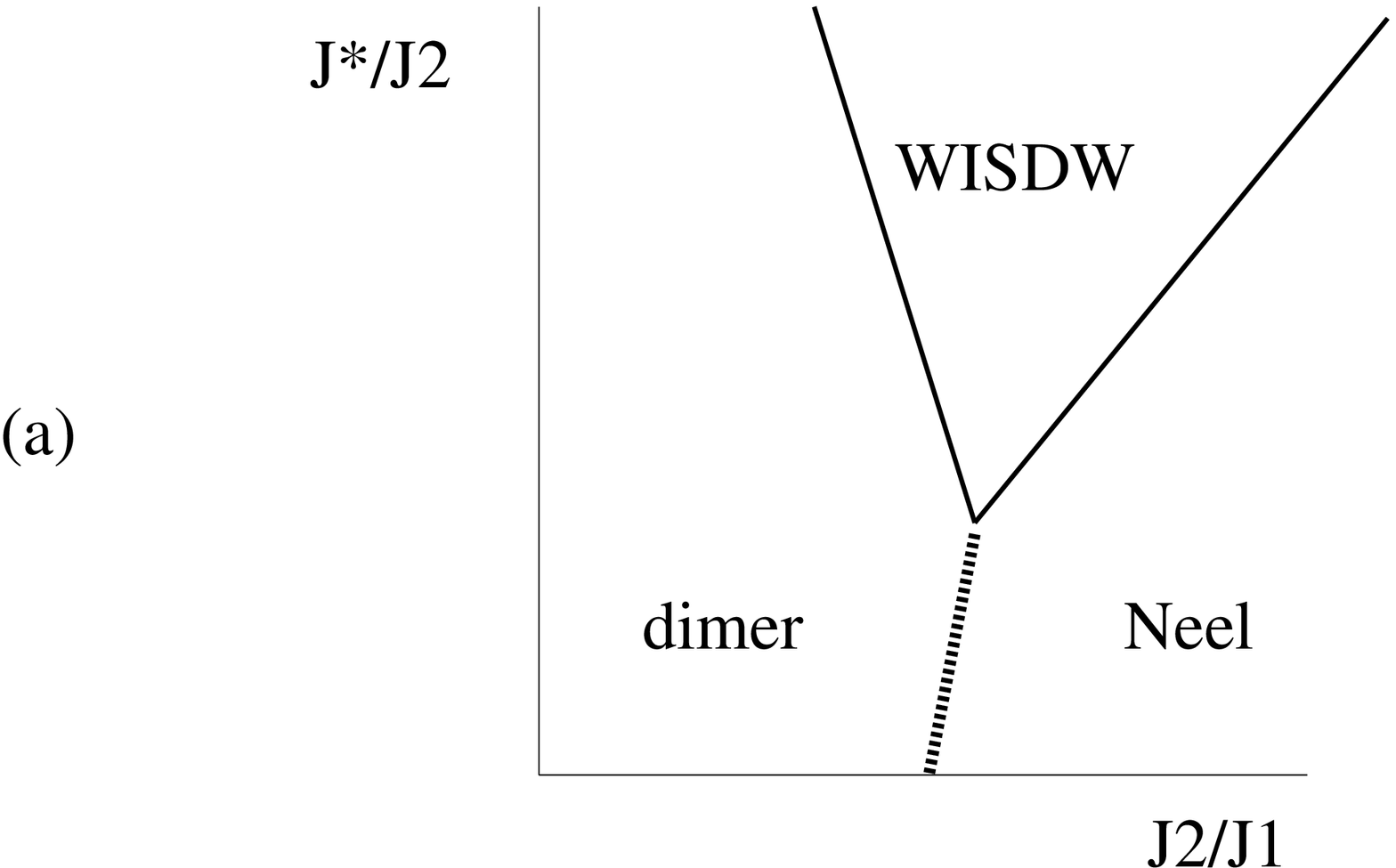}}
\psfrag{(b)}{\bf (b)}
\psfrag{M*}{$M^*$}
\psfrag{WISDW}{\bf WISDW}
\psfrag{Neel}{\bf N{\'e}el}
\psfrag{dimer} {\bf dimer}
\psfrag{J2/J1}{$J_2/J_1$}
\psfrag{J*/J2}{$J^*/J_2$}
\centerline{
\fig{6cm}{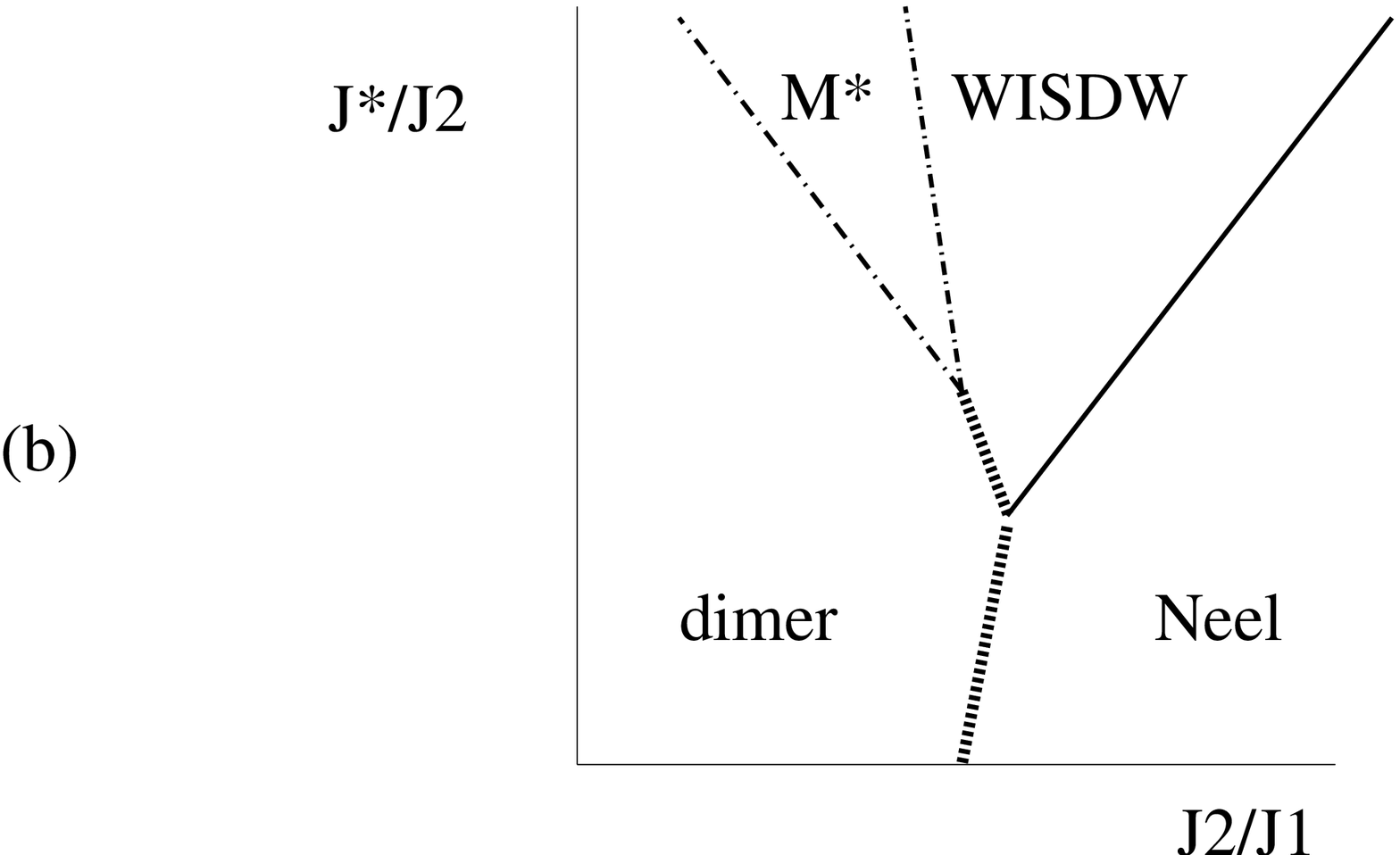}}
\caption{Schematic zero-temperature phase diagrams for GSS models upon 
  increasing a parameter, such as $J^*$, that favors dimer-boson
  condensation in (a) three dimensions, (b) two dimensions.  In
  three dimensions, only the dimer, N{\'e}el, and Weakly Incommensurate
  Spin Density Wave (WISDW) states need be present.  In two
  dimensions, fluctuation effects may open up a magnetically
  fractionalized region $M^*$ between the dimer and WISDW states.  Thick
  (hashed) lines indicate first-order transitions, thin solid lines
  second-order transitions, and dot-dashed lines transitions of
  unknown order.   }
\label{fig:schempds}
\end{figure}

Finally, we come back to the two-dimensional GSS models, and consider
the effect of critical fluctuations whose description is crucial at
the upper critical dimension. The natural framework to study these
fluctuations is provided by the renormalization group (RG) : in that
context, fluctuations correspond to either marginally irrelevant or
relevant operators in $d=2$.  We derive the one loop RG equations for
both the general DFT (bose-condensate transitions) and the field
theory describing the transition from the N{\'e}el state to the WISDW
phase.  For this latter transition, the 'gaussian' fixed point is
found to be stable for a range of parameters. This shows that
fluctuations do not destablize the transition suggested by the
mean-field behaviour, as shown in Fig.\ref{fig:schempds}.  For the
full DFT, which contains additional cubic terms allowed by the
existence of non-collinears order parameters, we have studied the RG
flow numerically and analytically.  All our approaches to the
analysis of the RG equations were consistent with a run-away flow of
the coupling constants.  This situation corresponds to marginally
relevant perturbations at the gaussian fixed point. This suggest a
different scenario in 2D than the MFT description of the 3D GSS models
(Fig.\ref{fig:schempds}).

It is tempting to connect the fluctuation-dominated regime near the
mean-field critical point with earlier theories of
``quantum-disordered'' non-collinear magnetic states.\cite{Subir}\ In
particular, it has been suggested that models with {\sl classically}
incommensurate non-collinear order may instead exhibit exotic
``fractionalized'' quantum paramagnets in the presence of strong
quantum fluctuations (e.g. for sufficiently small spin and/or
frustrated interactions).  This conclusion is based on analysis of
sigma models in which {\sl local} (in space and time) non-collinear
order is assumed, but that the orientation of this non-collinear order
parameter fluctuates quantum-mechanically.  The ``run-away'' flows in
the RG treatment suggest such a description might apply.  First, the RG
instability is driven by novel {\sl cubic interaction terms} that are
non-zero only for non-collinear spin configurations, and
present in the GSS model due to its unusual symmetries.  Further,
these run-away flows lead to a large fluctuation-induced ``critical
point shift'', so that the true critical point occurs deep within the
regime in which a local order-parameter amplitude is established.
A possible interpretation of our RG results
is therefore that the mean-field WISDW state in the vicinity of the critical
regime is replaced by such a fractionalized phase.  Such 
fractionalized states have a number of remarkable
properties\cite{Subir,fracts}\ 
(topological order, deconfined spin-$1/2$ excitations, etc.), which
make this an exciting possibility.  Notably, the presence of a
continuous quantum phase transition from the dimer state to a
fractionalized quantum paramagnet has been suggested recently by
Marston {\sl et. al.} using completely different large-$N$
methods.\cite{subir-private}
Our approach does not allow us to address the {\sl order} of such a
putative quantum critical point.  A schematic phase diagram for the
two-dimensional case is shown in Fig.~\ref{fig:schempds}b.  

The remainder of the paper is organized as follows.  In Sec.~II, we
introduce the dimer-boson representation and write down the general
DFT allowed for GSS models in coarse-grained variables.  We also show
how this general DFT reduces to a ``sub-model'' if an explicit
bond-alternation is added to the GSS model to break $90$-degree
rotational symmetry.  In Sec.~III, we discuss the structure of the
mean-field limit of the DFT, determining the full phase diagram for
the sub-model, and describing the types of ordering in the general
DFT.  We also prove that for the general DFT a generic Bose condensate
with local N{\'e}el order is unstable to a weakly incommensurate
spiral (WISDW).  In Sec.~IV, we show how a true antiferromagnet with
commensurate ($(\pi,\pi)$) N{\'e}el order and no other broken
symmetries is described as a Mott insulator, and determine the
effective field theory for the AF-WISDW transition.  Finally, in
Sec.~V we analyze the effect of quantum fluctuations in two dimensions
using the RG, both for the AF-WISDW and the dimer-WISDW critical
points (or more generally transitions out of the dimer state into Bose
condensates).  Details of the RG calculations are given in an
appendix.

\section{Models and bond-operator representation}
\label{sec:models}

\subsection{Generalized Shastry-Sutherland models and Dimer Field
theories}\label{sec:GSS-DFT}

In this paper, we explore the behavior of {\sl Generalized} SS (GSS)
models using a {\sl Dimer Field Theory} (DFT).  By GSS, we consider in
a general manner any model having the same symmetries as the SS
Hamiltonian (see below), preserving the exact ground state, and
without long-range 
(i.e. power-law) interactions.  The DFT of such models is obtained in
two stages.  First, the original spin model is rewritten in terms of
triplet bond operators.  As pointed out originally in
Ref.~\onlinecite{Sachdev90},\cite{comment}\ a pair of spins can be
represented {\sl exactly} by a triplet of hard-core bosons. Hence
\begin{subequations}\label{boson-spin}
\begin{eqnarray}
  \vec{S}_a({\bf x}) & = & \frac{1}{2} \left[
    \vec{b}_{1{\bf x}}^{\vphantom\dagger}
    + \vec{b}^\dagger_{1{\bf x}} + i \vec{b}^\dagger_{1{\bf x}}\times
    \vec{b}_{1{\bf x}}^{\vphantom\dagger}\right], \label{bopsfirst} \\
  \vec{S}_b({\bf x}) & = & \frac{1}{2} \left[
    - \vec{b}_{1{\bf x}}^{\vphantom\dagger}
    - \vec{b}^\dagger_{1{\bf x}} + i \vec{b}^\dagger_{1{\bf x}}\times
    \vec{b}_{1{\bf x}}^{\vphantom\dagger}\right], \label{bopssecond} \\
  \vec{S}_c({\bf x}) & = & \frac{1}{2} \left[
    \vec{b}_{2{\bf x}}^{\vphantom\dagger}
    + \vec{b}^\dagger_{2{\bf x}} + i \vec{b}^\dagger_{2{\bf x}}\times
    \vec{b}_{2{\bf x}}^{\vphantom\dagger}\right], \\
  \vec{S}_d({\bf x}) & = & \frac{1}{2} \left[
    - \vec{b}_{2{\bf x}}^{\vphantom\dagger}
    - \vec{b}^\dagger_{2{\bf x}} + i \vec{b}^\dagger_{2{\bf x}}\times
    \vec{b}_{2{\bf x}}^{\vphantom\dagger}\right], 
\end{eqnarray}
\end{subequations}
is an exact rewriting of the original spin operators in terms of
bosons, $[b_{i{\bf x}}^{\vphantom\dagger},b_{j{\bf x'}}^\dagger] =
\delta_{ij} \delta_{\bf xx'}$, provided the hard-core constraint,
\begin{equation}
\vec{b}^\dagger_{1{\bf x}}\cdot  \vec{b}^{\vphantom\dagger}_{1{\bf x}},
\vec{b}^\dagger_{2{\bf x}}\cdot  \vec{b}^{\vphantom\dagger}_{2{\bf x}} 
= 0,1, \label{bopslast}
\end{equation}
is enforced.  
A third usefull relation, that will be used later in this paper, can
be deduced from the above
definitions (\ref{boson-spin}) with the constraint (\ref{bopslast}). 
It  expresses the chirality on each bond : 
\begin{equation}\label{bopslastlast}
2~\vec{S}_a\times \vec{S}_{b}=i
(\vec{b}^{\dagger}_{1{\bf x}}-\vec{b}_{1{\bf x}}).    
\end{equation}
A general Hamiltonian for a GSS model can thus be
rewritten as a local model of interacting bosons via
Eqs.~\ref{boson-spin}-\ref{bopslast}.  This bond-operator
representation is particularly useful for GSS models in that it
captures the dimer eigenstate in the simplest possible way -- as the
boson vacuum.  The representation becomes awkward in the original SS
model when $J_2 \gg J_1$, in which the 2D square-lattice Heisenberg
model is approached.  This is because the canonical transformation in
Eqs.~\ref{boson-spin}-\ref{bopslast}, while exact, does not respect the
space-group symmetries of the square lattice.  Away from this limit,
however, all the {\sl physical} symmetries remain explicit.  

Using Eqs.~(\ref{boson-spin}-\ref{bopslast}), any interaction between
hard-core bosons can also be inverted and rewritten in terms of the
original spins.  It is particularly illuminating to rewrite the
four-spin interaction in Eq.~(\ref{eq:newints}),
\begin{equation}
  H^* = -J^* \sum_{\langle{\bf xx'}\rangle} \sum_{i=1,2} 
\left( \vec{b}_{i{\bf x}}^\dagger \cdot 
       \vec{b}_{i{\bf x}'}^{\vphantom\dagger} + {\rm
    h.c.}\right), 
\end{equation}
which becomes a simple boson hopping term.  Formally, although we do
not explore this route in detail, $J^*$ introduced in this manner can
act as a control parameter for the dimer boson theory.  In particular, 
as $J^*$ is increased, the boson kinetic energy becomes dominant, and
the system is better and better approximated as a weakly-interacting
bose gas.  For such a gas, the bose condensate approximation to the
interacting ground state is obviously very good.

Next, we adopt a {\sl coarse-grained} effective field theory point
of view.  In particular, we will assume that any ordering and/or
low-energy fluctuations in the system at most weakly (i.e. at very
long wavelengths) breaks lattice translational symmetry.  By
translational symmetry, we mean translation by a Bravais lattice
vector ${\bf x} = n_1 {\bf a}_1 + n_2 {\bf a}_2$, with integer
$n_1,n_2$ and ${\bf a}_{1,2}$ indicated in Fig.~\ref{fig:ss1}.  Note that the
because of the rather large four-site unit cell, this requirement is
not very restrictive.  In particular, both the dimer and N{\'e}el order 
are translationally invariant in this strict sense.

With this assumption, we can imagine formally integrating out (in the
path-integral sense) the degrees of freedom (Fourier modes of the
$b_i^{\vphantom\dagger},b_i^\dagger$ operators) with wavevectors
$|{\bf q}|>\Lambda$, where $\Lambda \ll \pi$ is a cut-off defining a sphere
around the origin in momentum space.  This procedure defines an
(approximate) continuum limit, $\vec{b}_{i{\bf x}} \rightarrow
\vec{\phi}_i({\bf x})$ , $\vec{b}^\dagger_{i{\bf x}} \rightarrow
\vec{\phi}^*_i({\bf x})$, with a pair of continuum triplet fields
$\vec{\phi}^{\vphantom\dagger}_i,\vec{\phi}_i^\dagger$.  Because these
continuum fields are effectively averages of the microscopic bosons
over many unit cells, this coarse-graining relaxes the
hard-core constraint.  Thus we arrive at an effective ``soft spin''
field theory.  

In practice, for a general GSS Hamiltonian, it is impossible to
integrate out the short-wavelength modes explicitly.  Instead, in the
spirit of classical Landau theory, we use the constraints of symmetry
and the presence of an exact dimer ground state to determine the form
of the long-wavelength effective action in an expansion in powers of
the fields and in space-time gradients.  
 Beside the invariance by time-reversal, the symmetries of the initial
SS model (and by extension of the GSS models we consider) can be
determined by inspection of the geometry of the corresponding lattice
(Fig.~\ref{fig:ss1}) : 
\begin{itemize}

\item Time reversal :
$\vec{S}\rightarrow -\vec{S}~;~\vec{\phi}_{a}\rightarrow -\vec{\phi}_{a}^{*}
~;~\tau \rightarrow -\tau   $

\item $\pi/2$  rotation :
 $\vec{\phi}_{1}\rightarrow \vec{\phi}_{2}; \vec{\phi}_{2}\rightarrow 
-\vec{\phi}_{1}
;x\rightarrow y; y\rightarrow -x$

\item Reflection with respect to $x$ :
$x\rightarrow -x ~;~ \vec{\phi}_{1}\rightarrow -\vec{\phi}_{1}$

\item Reflection with respect to $y$ :
$y\rightarrow -y ~;~ \vec{\phi}_{2}\rightarrow -\vec{\phi}_{2}$

\end{itemize}

We keep terms up to fourth
order in the fields and second order in spatial gradients.  Breaking
up the effective action by powers of the fields gives $S=
S^{(2)} + S^{(3)} + S^{(4)}$.  
\startlargeeq

The quadratic terms are
\begin{subequations}
\label{def-s}
\begin{equation}
S^{(2)} =  \int_{{\bf x}, \tau } \Bigg\{
\vec{\phi}_{1}^{*}
\left[Z~ \partial_{\tau} + r -
  c_{1}\partial_{x}^{2}-c_{2}\partial_{y}^{2} \right] 
\vec{\phi}_{1} +\vec{\phi}_{2}^{*}
\left[Z~ \partial_{\tau}+ r
  -c_{2}\partial_{x}^{2}-c_{1}\partial_{y}^{2} \right] 
\vec{\phi}_{2}  - d \left( 
\vec{\phi}_{1}^{*}\partial_{x}\partial_{y}\vec{\phi}_{2}
          +\vec{\phi}_{2}^{*}\partial_{x}\partial_{y}\vec{\phi}_{1}\right)
        \Bigg\}. \label{s2}
\end{equation}
The cubic terms may be written
\begin{equation}
S^{(3)} = 
i ~\int_{{\bf x}\tau}
\bigg\{  
\bigg[ 
g_{0}
\vec{\phi}_{1}^{*} \cdot
\partial_{x} \vec{\phi}_{1}\times \vec{\phi}_{1}
+ g_1 \vec{\phi}_{2}^{*} \cdot
\partial_{x} \vec{\phi}_{1}\times \vec{\phi}_{2}
+ g_{2}
\vec{\phi}_{1}^{*} \cdot
 \partial_{x}\vec{\phi}_{2}\times\vec{\phi}_{2}
+ g_{4}
\vec{\phi}_{2}^{*} \cdot
\partial_{x} \vec{\phi}_{2}\times \vec{\phi}_{1}
- {\rm h.c.} \bigg]
+ [ y\leftrightarrow x,
1\leftrightarrow 2] \bigg\} , \label{s3}
\end{equation}
where the product ($\times$) indicates the (vector)cross-product.
Finally, the allowed quartic terms are
\begin{eqnarray}
S^{(4)} & = & 
\int_{{\bf x}\tau}
\biggl\{ 
 u_{1}\left[ 
(\vec{\phi}_{1}^{*}.\vec{\phi}_{1})(\vec{\phi}_{1}^{*}.\vec{\phi}_{1})
            +1\leftrightarrow 2 \right]
+2u_{2}\left[(\vec{\phi}_{1}.\vec{\phi}_{1}^{*})(\vec{\phi}_{2}.\vec{\phi}_{2}^{*}) 
\right]
+u_{3}\left[(\vec{\phi}_{1}.\vec{\phi}_{2}^{*})(\vec{\phi}_{1}.\vec{\phi}_{2}^{*})
            +1\leftrightarrow 2 \right]
+2 
u_{4}\left[(\vec{\phi}_{1}.\vec{\phi}_{2}^{*})(\vec{\phi}_{2}.\vec{\phi}_{1}^{*})\right]
\nonumber \\
& & 
+
 v_{1}\left[(\vec{\phi}_{1}.\vec{\phi}_{1})(\vec{\phi}_{1}^{*}.\vec{\phi}_{1}^{
*})
            +1\leftrightarrow 2 \right]
+v_{2}\left[(\vec{\phi}_{2}.\vec{\phi}_{2})(\vec{\phi}_{1}^{*}.\vec{\phi}_{1}^{*})
            +1\leftrightarrow 2 \right]
+4 
v_{3}\left[(\vec{\phi}_{1}.\vec{\phi}_{2})(\vec{\phi}_{1}^{*}.\vec{\phi}_{2}^{*})\right]
\nonumber 
\\
& & 
+
 w_{1}\left[(\vec{\phi}_{1}.\vec{\phi}_{1})(\vec{\phi}_{1}.\vec{\phi}_{1}^{*})
            + 1\leftrightarrow 2  \right]
+w_{2}\left[(\vec{\phi}_{2}.\vec{\phi}_{2})(\vec{\phi}_{1}.\vec{\phi}_{1}^{*})
            + 1\leftrightarrow 2  \right]\
+4 
w_{3}\left[(\vec{\phi}_{1}.\vec{\phi}_{2})(\vec{\phi}_{2}.\vec{\phi}_{1}^{*})
            + 1\leftrightarrow 2  \right]
+h.c. \biggr\}.  
\label{s4}
\end{eqnarray}
\end{subequations}
\stoplargeeq
Eqs.~\ref{def-s}\ embody numerous unique features of the GSS
models.  For $r>0$, the quadratic action $S^{(2)}$ in Eq.~\ref{s2}\
indicates the presence of a gap of order $r$ for triplet excitations.
For $r$ sufficiently positive, the ground state of the system is the
boson vacuum, corresponding to the exact dimer eigenstate.  Neglecting 
for the moment the cubic and quartic terms, as $r\rightarrow 0$ the
gap for triplet excitations vanishes.  At this (mean-field)
critical point ($r=0$), the space-time scaling is highly anisotropic:
naively, $\omega \sim k^z$ with the dynamical exponent $z=2$.  This
strongly anisotropic scaling is very different from the $z=1$ behavior 
expected at a generic $O(3)$ transition described by, e.g. the
non-linear sigma model.  This anisotropy is due to the
fluctuationless nature of the dimer state and consequent reduction of
``quantum fluctuations'' near the putative critical point.  

A further constraint on Eqs.~\ref{def-s}\ due to the presence of
the exact dimer state is the absence of ``anomalous'' terms involving
products {\sl only} of creation or annihilation operators.  For
instance, consider the introduction of an exchange coupling between
$a$ and $b$ spins in {\sl different} unit cells, which
destroys the exactness of the dimer state.  This interaction leads
directly to an anomalous term of the form $J^{''} 
(\vec{\phi}_i\cdot\vec{\phi}_i +
{\rm c.c.})$, which induces a 
cross-over to more conventional $z=1$ behavior.  

An unusual feature of the above effective action is the cubic terms in
Eq.~\ref{s3}.  The presence of such cubic invariants involving single
spatial gradients is a unique feature of the GSS models.  
They are allowed by symmetry due to a combination of factors: first, the
complex nature of the triplet $\vec{\phi}_i$ fields,
(corresponding to non-linear order parameters in the vertical and
horizontal directions)  makes it possible
to construct a non-vanishing triple-product; 
and second, the two
inequivalent dimers per unit cell give rise to the ``flavor'' index
$i=1,2$ which in fact transforms like a {\sl spatial} vector index
under discrete lattice point group operations, allowing a linear
gradient to complete a point-group scalar.
 Note that even the sub-model defined in the next section, obtained by
forgetting {\it e.g} the order parameters on the horizontal bonds, contains 
such a  cubic term : the presence of the two complex vector
fields $\vec{\phi}_{1}$ and $\vec{\phi}_{2}$ only
enlarges the number of independent cubic terms, but does not change
the essential features.  We will see that these cubic terms
{\sl qualitatively} modify the properties of these field theories.
Since such terms are non-vanishing {\sl only} for field configurations 
in which the fluctuating magnetization has multiple orthogonal components, this
indicates that {\sl non-collinear} magnetic order plays a significant
role in the physics.

\subsection{Sub-Model}\label{sec:sub-model}

Even in 3D, and certainly in 2D, the non-quadratic terms in the
effective action are crucial in determining the nature of the ordered
phase which occurs for $r<0$.  To get a feeling for the effects of the 
cubic and quartic interactions, it is helpful to consider a
deformation of the GSS model obtained by increasing the exchange
constant along the solid vertical bonds, leaving the solid horizontal
bonds unchanged.  This splits the degeneracy of the two branches of triplet
excitations, and has the effect in the field theory of adding an
additional $\delta J~ \vec{\phi}_2^*  \cdot \vec{\phi}_2$ term to $S^{(2)}$, 
thereby creating an effective quadratic Landau coefficient 
$r_2 = r+\delta J > r$ for the $\vec{\phi}_2$ field.  Thus when
$r\rightarrow 0$, the coefficient $r_2$ remains positive, and the
still massive $\vec{\phi}_2$ field can be integrated out in the critical 
region.  One is left with a ``sub-model'' involving only terms (with
slightly renormalized coefficients) in Eqs.~\ref{def-s} 
depending only on the field $\vec{\phi}_1$:  
\startlargeeq
\begin{multline}\label{eq:sm}
  S_{\rm sm}=\int_{{\bf x}\tau} \bigg\{
  \vec{\phi}^{*}(Z\partial_{\tau}+ R - C{\bf \nabla}^{2})\vec{\phi}  
  -iG\partial_{x}(\vec{\phi}+\vec{\phi}^{*})\cdot\vec{\phi}\times 
\vec{\phi}^{*} \\
  +U(\vec{\phi} \cdot\vec{\phi}^{*})^{2}+V(\vec{\phi} \cdot\vec{\phi} 
)(\vec{\phi}^{*}\cdot\vec{\phi}^{*})
  +W\left[ (\vec{\phi} \cdot\vec{\phi} 
)(\vec{\phi}\cdot\vec{\phi}^{*})+c.c.\right] \bigg\},
\end{multline}
\stoplargeeq
\noindent where we have, without loss of generality, rescaled the spatial axes
to make the kinetic term (with coefficient $C$) isotropic (note that
this is not possible in the full model).  From the point of view of
symmetries, this 'sub-model' can be defined as the DFT invariant under
the above defined time reversal symmetry, reflection with respect to
$x$ and $y$ (the latter having a trivial effect on the field :
$\phi\to \phi$), and for which the boson vacuum is an exact
eigenstate.  The mean-field theory of Eq.~\ref{eq:sm}\ is solved
exactly in Sec.~\ref{sec:smmft}.

\section{Landau Theory and Mean-Field Phase Diagram}
\label{sec:landau}

Remarkably, simple power counting (see Sec.~\ref{sec:RG}) indicates
that {\sl both} the cubic and quartic terms (in $S^{(3)}$ and
$S^{(4)}$ respectively) are {\sl marginal} in the Renormalization
Group (RG) sense in two spatial dimensions (2D), which plays the role
of the upper critical dimension in critical phenomena.  In three
dimensions, the cubic and quartic interactions are {\sl irrelevant} at
the critical point, affording the possibility of an unusual mean-field
critical point in a three-dimensional (3D) system.  In this section,
we study this mean-field behaviour in details. For pedagogical
reasons, we will not consider the MFT of the full model (\ref{def-s})
as the physics of the solutions may be obscured by the intrinsic
complexity of this field theory. Instead, we will focus on the
above-defined sub-model whose mean-field analysis will capture the
essential physics of the full MFT phase space. This analysis, carried
in the next section, establishes the existence of three phases : a
phase where spins are antiparallel on each dimer, a chiral phase in
which there is no static local magnetization, but the two spins
maintain a fixed normal relative orientation while ``rotating''
quantum-mechanically with respect with each other on the dimer, and a
spiral phase. The discussion of the antiferromagnetic ``N{\'e}el''
phase of the GSS model, which corresponds to a Mott insulator in terms
of the lattice bosons, is postponed to the next section.

\subsection{Sub-Model}
\subsubsection{Mean-Field solutions}
\label{sec:smmft}

Within the sub-model, consider first $G=0$, in which the action
becomes rotationally-invariant (under spatial rotations).  If,
furthermore, $V=W=0$, Eq.~\ref{eq:sm}\ has an $O(6)$ symmetry under
global orthogonal rotations of the six-component vector composed of
the real and imaginary parts of $\vec{\phi}$.  Thus, for $V=W=0$, in
mean-field theory (MFT), the system undergoes a continuous $O(6)$
symmetry-breaking transition as $R \rightarrow 0$ to a state with an
arbitrary complex expectation value $\langle \vec{\phi} \rangle \neq 0$, 
and $\langle \vec{\phi} \rangle^* \cdot \langle \vec{\phi} \rangle = -
R/2U$ for $R<0$.  The $V$ and $W$ terms reduce the $O(6)$ symmetry
down to $O(3)\times U(1)$ and $O(3)$, respectively.  For $V>0$, it is
helpful to decompose $\vec{\phi}$ into real and imaginary parts:
$\vec{\phi} = \vec{\eta} + i \vec{\xi}$, where $\vec{\eta}$ and $\vec{\xi}$ are
real vectors.  For $W=0$ and $V>0$, the lowest energy states then have 
$|\vec{\eta}|^2=|\vec{\xi}|^2$ and $\vec{\eta}\cdot\vec{\xi} = 0$, so
that the order 
parameter becomes an orthogonal pair of fixed length vectors
comprising a ``frame'', similar to the order parameter in a bi-axial
nematic liquid crystal.  Including a non-zero $W$ (with still $V>0$)
favors unequal (but still orthogonal) real
and imaginary parts, and the MFT minima becomes instead $|\vec{\eta}| =
|\vec{\phi}| \cos\psi, |\vec{\xi}|=|\vec{\phi}|\sin\psi$, with $\cos 2\psi = -
W/V$.  On the other hand, if $V<0$ and $W=0$, the MFT ground state is
of the form $\langle \vec{\phi}\rangle = |\vec{\phi}| \hat{ n} e^{i\alpha}$,
with an arbitrary real unit vector $\hat{ n}$ and phase 
$\alpha$, 
corresponding to parallel real and imaginary components.  Including
non-zero $W$ simply breaks the phase degeneracy to favor
$e^{i\alpha}=1,i$ for $W<0$ and $W>0$, respectively.

Now consider including $G\neq 0$.  It is instructive to rewrite the
cubic term as $-iG\partial_{x}(\vec{\phi}+\vec{\phi}^{*})\cdot\vec{\phi}\times 
\vec{\phi}^{*}
= 4 G \partial_x \vec{\eta} \cdot \vec{\eta}\times \vec{\xi}$.  By
inspection, a non-zero $G$ thus favors {\sl spatially non-uniform}
configurations in which $\vec{\eta}$ {\sl precesses} around $\vec{\xi}$ as
one proceeds along the $x$ axis.  Physically (see below), this
precession corresponds to spiral magnetic order.  We allow for this by
considering $x$-dependent configurations.  Defining polar and
azimuthal angles, $\theta$ and $\varphi$ respectively, of $\vec{\eta}$
in the spherical coordinates defined with $\vec{\xi}$ along the polar
axis, this term is written $-4G|\vec{\phi}|^3 \cos^2\psi\sin\psi
\sin^2\theta \partial_x \varphi$.  For fixed, constant $|\vec{\phi}|$ and
$\theta$, the mean-field Lagrange density becomes then
\startlargeeq
\begin{multline}
  {\cal L}_{\rm sm}^{\rm MF} = C |\vec{\phi}|^2 \cos^2 \psi \sin^2 \theta
    (\partial_x\varphi)^2 - 4 G |\vec{\phi}|^3
  \cos^2\psi\sin\psi\sin^2\theta \partial_x\varphi \\
+ |\vec{\phi}|^4 \bigg\{U+
  V\left[\cos^2 2\psi  + \sin^2 2\psi \cos^2\theta\right] + 2 W \cos
  2\psi \bigg\}.
\end{multline}
The optimal precession wavevector is thus $Q = \partial_x
\varphi = 2 (G/C) |\vec{\phi}| \sin\psi/\sin\theta$ (the
singularity as $\theta\rightarrow 0$ does not influence the results).
Using this frequency, the Lagrange density becomes
\begin{equation}
  {\cal L}_{\rm sm}^{\rm MF} = |\vec{\phi}|^4 \bigg\{ - (G^2/C) 
  \sin^2 2\psi\sin^2\theta + U+
  V\left[\cos^2 2\psi  + \sin^2 2\psi \cos^2\theta\right] + 2 W \cos
  2\psi \bigg\}.
\end{equation}
\stoplargeeq
\noindent Note that the contribution arising from
minimization over $Q$ is of the same order ($O(|\vec{\phi}|^4)$) as the $U$
and $V$ interactions, which is the MFT manifestation of $G$ being
marginal at the upper critical dimension.
The minimum energy (action) configurations are then found to be 
\begin{align}
(i)  & \quad \theta = \pi/2,\quad 
\cos 2\psi = - \frac{W}{V+G^2/C} ,\nonumber\\  
&\qquad \qquad \qquad  V+
  G^2/C > |W| > 0, \nonumber \\
(ii)  & \quad\theta = \pi/2,\psi = \frac{\pi}{2} \Theta(W)   \qquad  
   |W|> V+ G^2/C  > 0, \nonumber \\
(iii)  & \quad\theta = 0, \psi = \frac{\pi}{2} \Theta(W) 
 \qquad\qquad  V + G^2/C < 0 ,\label{MFsols}
\end{align}
where $\Theta(W)$ is the Heavyside step function.  In fact, the latter 
two solutions are physically equivalent, since when $\psi=0$ or
$\psi=\pi/2$, $\vec{\phi}$ is either pure real or pure imaginary, and
$\theta$ has no physical significance.  Thus the precessing solutions
occur only for $V+G^2/C > |W| >0$.  In this regime, however, a
continuous transition occurs within mean-field theory (provided the
stability condition $U> G^2/C + W^2/(V+G^2/C)$ holds) from the dimer
to a spiral phase.  

\subsubsection{Phases}
\label{sec:phasediag}

The physical 
meaning of these phases is best described by the three observables,
\begin{eqnarray}
  \vec{N}_{1{\bf x}} \equiv \vec{S}_a({\bf x}) +
  \vec{S}_b({\bf x}) & = & i 
  \vec{b}_{1{\bf x}}^\dagger \times \vec{b}_{1{\bf x}} ,
  \label{eq:stot} \\
  \Delta\vec{S}_{1{\bf x}} \equiv \vec{S}_a({\bf x}) - \vec{S}_b({\bf x}) & = &
  \vec{b}_{1{\bf x}}^{\vphantom\dagger} +   \vec{b}_{1{\bf x}}^\dagger
  , \label{eq:deltas} \\
  \vec{T}_1({\bf x}) \equiv 2\vec{S}_a({\bf x}) \times \vec{S}_b({\bf
    x}) & = & - 
  i(\vec{b}_{1{\bf x}}^{\vphantom\dagger} - \vec{b}_{1{\bf
      x}}^\dagger). \label{eq:chirality}
\end{eqnarray}
 where $\vec{T} $ stands for the chirality defined on each bond. 
The relation to the boson variables can be shown from
Eqs.~\ref{bopsfirst}-\ref{bopssecond} and the hard-core condition.

We expect that, using the above decomposition of $\vec{\phi}$ into real and
imaginary parts,   
$\langle \Delta 
\vec{S}_1 \rangle = 2{\rm Re}\langle \vec{\phi}_1\rangle = 2\vec{\eta}$
and $\langle \vec{T}_1 \rangle = 2 \vec{\xi}$.  If {\sl both} $\vec{\eta}$ and
$\vec{\xi}$ are non-zero, then general principles require
that $\langle \vec{N}_1\rangle \neq 0$, and we expect
$\langle \vec{N}_1\rangle \propto \vec{\eta} \times \vec{\xi}$.
The converse, however, does not follow: it is possible that
$\langle\vec{S}_1^{\rm tot}\rangle = \langle i\vec{b}_1^\dagger \times
\vec{b}_1 \rangle \neq 0$ but $\langle \vec{b}_1\rangle = 0$ (though
this is not a mean-field state).

Armed with these relations, we can caracterize the three different
mean-field phases defined by (\ref{MFsols}) :

(i) for $V+G^2/C>|W|>0$ have all three ``order parameters''
non-vanishing.  Since $\vec{\xi}$ is constant in these solutions, the
``chirality'' $\vec{T}_1$ is fixed on each horizontal bond, as
$\vec{\eta}$ precesses, both $\vec{N}_1$ and $\Delta
\vec{S}_1$ spiral from bond to bond along the $x$ direction.  The
non-spiral solutions are of two varieties. Thus in
this {\sl spiral} phase all three
quantities are non-vanishing, with a fixed (i.e. spatially constant)
chirality $\vec{T}_1$ on each horizontal bond.  The spins themselves,
however, spiral from bond to bond along the $x$ direction, maintaining
the condition $\langle\vec{N}_1 \rangle \cdot \langle \Delta
\vec{S}_1 \rangle = 0$.  In the special limit $G \rightarrow 0$, the {\sl
  pitch}  of this spiral vanishes, and the spiral phase goes over to a 
sort of biaxial spin nematic. 

(ii) for $W>0,V+G^2/C$, only
$\vec{\xi} \neq 0$, and one has a {\sl chiral} phase: $\vec{T}_1 \propto
\vec{\xi} \neq 0$  but the spins themselves are disordered, i.e. $\langle
\vec{S}_a\rangle = \langle \vec{S}_b\rangle = 0$.  

(iii) for
$W<0,-|V+G^2/C|$, the chirality vanishes and the two spins on each
bond are antiparallel, $\langle \vec{S}_a \rangle = - \langle
\vec{S}_b \rangle \neq 0$.  This {\sl antiparallel} phase is thus
characterized by a single non vanishing order parameter 
 $\langle \Delta \vec{S}_1 \rangle
\neq 0$,

 In MFT, continuous phase transitions are
possible between all pairs amongst the four (antiparallel, chiral,
spiral, and dimer) phases except between the anti-parallel and chiral
states, which is a first-order transition.
A schematic cut through the mean-field
phase diagram for $R<0$ is shown in Fig.~\ref{fig:smpd}.

\begin{figure}[htb]
\psfrag{Antiparallel}{{\Large Antiparallel}}
\psfrag{Spiral}{{\Large Spiral}}
\psfrag{Chiral}{{\Large Chiral}}
\psfrag{W}{$W$}\psfrag{2}{}\psfrag{/C}{}\psfrag{V+G}{$V+G^{2}/C$}
\hspace{0.2in}\fig{2.5in}{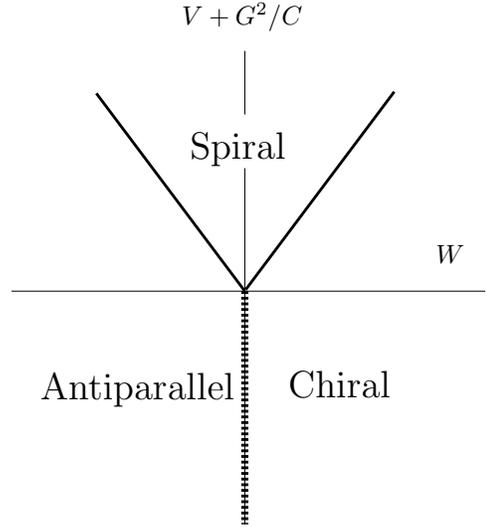}
\caption{Schematic cut through the mean-field phase diagram of the
  sub-model for $R<0$.}
\label{fig:smpd}
\end{figure}

Conspicuously absent from the MFT is a non-chiral
($\langle\vec{T}_1\rangle =0$) state with spontaneous magnetization on
each dimer ($\langle \vec{N}_1\rangle \neq 0$) but equal
moment for each of the two dimer spins ($\langle\Delta\vec{S}_1\rangle
= 0$).  As noted in the introduction, this is because such a state is
{\sl not} a 
triplet-boson condensate, $\langle \vec{b}_1\rangle=0$, i.e. not a
``superfluid''.  The absence of this ``N{\'e}el'' state is all
the more troubling insofar as it is this state which corresponds to
the antiferromagnetic state of the original SS model (with the
vertical bonds removed by the additional interactions of the
sub-model).  As discussed in Sec.~\ref{sec:mott}, such a ``N{\'e}el''
state is described in the bond-boson variables as a bosonic {\sl Mott
  insulator}.  Because such Mott insulators rely upon commensurability
of the boson number with the lattice (i.e. in this case there is
precisely one pair of aligned spins per horizontal bond), they are in
fact outside the simple continuum description of
Eqs.~\ref{def-s}\ or Eq.~\ref{eq:sm}.  A further consequence of
the identification of the ``N{\'e}el'' state with the Mott insulator
is that it cannot be obtained via a single continuous transition from
the dimer phase (see below). 
  
\subsection{Full DFT}
\subsubsection{Mean-Field Phases}

As emphasized above, the complexity of the full DFT renders an
exhaustive analysis of even its mean-field phase diagram less than
illuminating.  However, the general sorts of phases which can arise
are easily understood by analogy to the sub-model analysis above.  In
particular, with the two fields $\vec{\phi}_1$ and $\vec{\phi}_2$, one
expects ``direct products'' of the states obtained therein, i.e. with
real, imaginary, or non-collinear complex values independently for
each field.  Further, the quartic terms coupling both fields
($u_{2,3,4}$, $v_{2,3}$, and $w_{2,3,4}$) will select definite
relative orientations and phases of these order parameters.  Thus,
even without the possibility of spatially-varying (i.e. spiral)
solutions, the full DFT clearly sustains a panoply of unusual phases.

Much of this complexity, however, arises in regions of the phase
diagram in which we are uninterested.  For models not too much
deformed from the original Shastry-Sutherland hamiltonian, we expect a
strong tendency for ordered state to local antiferromagnetic
magnetization.  As shown above, such local order
($\vec{N}_1=-\vec{N}_2\neq 0$) requires non-collinear complex values
for both $\vec{\phi}_1$ and $\vec{\phi}_2$, and moreover a specific
orientation and relative phase to these two fields.  It is
straightforward to construct regions in the space of quartic couplings
in which such states are favored\cite{unpub}\ (although the task of
computing the {\sl full} such region is formidable!).  Indeed, an
effective Lagrangian in this subspace is naturally obtained by a
direct rewriting of the Shastry-Sutherland spin Hamiltonian using
Eqs.~(\ref{boson-spin}) and very naive coarse-graining of the quartic
interactions.\cite{unpub}\ Thus, in the absence of cubic terms, the
DFT could naturally describe transitions from the dimer state to a
state with N{\'e}el order, at least at the mean-field level.  As we show
below, however, the non-vanishing cubic terms modify this conclusion
significantly.

\subsubsection{Spiral instability of N{\'e}el order}

As noted above, a striking
 feature of the mean-field phase diagram for the sub-model
(Fig.~\ref{fig:smpd}) is the absence of a phase with commensurate
N{\'e}el order, i.e. with $\langle \vec{S}^{\rm tot}_{1{\bf x}}\rangle =
\vec{N}$ equal to a constant.  Instead, the only phase with a net
average on each bond moment is the spiral state, in which the spin
density wave order shifts to a small non-zero wavevector and precesses
in spin space.  Uniform N{\'e}el ordering occurs only in the limiting case in
which the quantum critical points to the dimer or anti-parallel states
are approached.

In this subsection, we show that this behavior continues to hold in
the full set of GSS models.  More precisely, without fine tuning of
parameters, in any 
{\sl  dimer bose condensate with a non-zero local N{\'e}el
order}, this N{\'e}el vector forms an incommensurate spiral with a long
pitch.  
 
 We consider a N{\'e}el phase corresponding
to uniform expectations values 
$\langle \vec{\phi}_{1}\rangle =A\hat{ e}_{1}+iB\hat{ e}_{2}$ and 
$\langle \vec{\phi}_{2}\rangle =\langle \vec{\phi}_{1}^{*}\rangle
  =A\hat{ e}_{1}-iB\hat{ e}_{2}$ where $\hat{ e}_{1}$ and $\hat{ e}_{2}$ are 
two
orthogonal vectors. 
 This state is caracterised as needed by antiferromagnetic order  
$\langle \vec{N}_{1}\rangle=i\langle \vec{\phi}_{1}^{*}\times 
\vec{\phi}_{1}\rangle
=-\langle \vec{N}_{2}\rangle=i\langle \vec{\phi}_{2}^{*}\times
\vec{\phi}_{2}\rangle =-2AB \hat{ e}_{3}$ where 
$\hat{ e}_{3}=\hat{ e}_{1}\times \hat{ e}_{3}$
 Moreover, on each bond the chirality is given by
$\vec{T}_{1}=2B\hat{ e}_{2}=-\vec{T}_{2}$, and 
$\Delta \vec{S}_{1}=\Delta \vec{S}_{2}=2A\hat{ e}_{1}$.

To study the stability of this N{\'e}el phase with respect to spiral
order, we consider spatial-dependant small rotations of the
normalized triad $\hat{ e}_{1},\hat{ e}_{2},\hat{ e}_{3}$
defined above. The rotations with respect to these three axis are
respectly parametrized by the three angles
$\theta_{1},\theta_{2},\theta_{3}$.  The usual gradient contribution to
the energy of these fluctuations, ${\cal L}_\theta^0$,  from the quatratic part
$S^{(2)}$ of the action (\ref{s2}) is :
\begin{multline}
{\cal L}_{\theta}^{0} = c_{1}\left[
B^{2}(\partial_{x}\theta_{1})^{2} 
+A^{2}(\partial_{x}\theta_{3})^{2}
+(A^{2}+B^{2})(\partial_{x}\theta_{3})^{2}
\right]\\
+(x\to y,c_{1}\to c_{2}).
\end{multline}

Next consider the effect of the cubic terms which depend explicitly on
the spatial variations of
$\vec{\phi}_{1},\vec{\phi}_{2}$ and are thus directly
sensitive to these chiral instabilities.  Note that
$\vec{\phi}_{1}=\vec{\phi}_{2}^{*}$ (needed to make
$\vec{N}_1=-\vec{N}_2$), and moreover the fluctuations considered
leave this equality unchanged. Hence the $g_{2}$ and $g_{4}$ terms in
(\ref{s3}) do not contribute to the energy of these local rotations.
{}From the two others, we get a quadratic contribution,
\begin{align}
{\cal L}_\theta^{1} = 4 (g_{0}-g_{1}) A^{2}B~ 
\theta_{1}(\partial_{x}\theta_{3}-\partial_{y}\theta_{3}). \label{linear}
\end{align}

The resulting action for the $\theta$ fluctuations is obviously not
definite positive, except with the special fine tuning $g_{0}=g_{1}$
or $A=0$ or $B=0$.  This completes the proof of the above assertion
that the ``N{\'e}el'' phase is unstable to a weak spiral deformation,
as a non-vanishing texture in $\theta$ will always develop for some
momentum ${\bf Q}$.  Although this proof was made specifically within
the DFT theory, the conclusion is in fact more general and rests
only upon the symmetries of the GSS models.  Briefly, the massless of
all three of the $\theta$ modes is required by Goldstone's theorem,
since there are no unbroken subgroups of the SU(2) spin-rotation
symmetry.    Similarly, linear gradient terms of the form in
Eq.~(\ref{linear}) are generically present by symmetry, since $\Delta
\vec{S}_{1{\bf x}}=\Delta\vec{S}_{2{\bf x}}\propto {\bf \hat{e}}_1$
breaks reflection invariance.  Thus the dimer bose condensate does not 
generically describe a state with uniform N{\'e}el order.

\section{Bosonic Mott Insulators}
\label{sec:mott}

\subsection{Sub-model}

As discussed in the introduction, it is clear from
Eqs.~\ref{eq:stot}-\ref{eq:chirality}\ that a state with spontaneous
collinear and aligned moments on the two sites of each bond of the SS
lattice cannot be described as a dimer-boson condensate.  Instead,
these are Mott insulators.  To understand the physics of such states,
we first discuss the simple example of the ``antiferromagnetic''
state in the bond-boson field theory for the sub-model.  
The only known non-superfluid ground states of interacting
boson systems without disorder are bose solids or Mott insulators, the
latter being most simply understood as a bose solid pinned by a
commensurate lattice potential.  Because the desired state does not
break translational symmetry, we are led to consider a bosonic Mott
insulator as a candidate state.  To do so, we are required to include
the effects of the underlying lattice.  A ``microscopic'' means of
including these lattice effects would be to return to the description
in terms of the $\vec{b}_{1{\bf x}}^{\vphantom\dagger}, \vec{b}_{1{\bf
    x}}^\dagger$ operators.  Instead, we continue to employ an {\sl
  effective} field theory approach, both for consistency with the
remainder of the paper and on the grounds that direct analytic
treatments of the microscopic model have no control parameter and are
hence unreliable.  On symmetry grounds, lattice effects can be
included by addition additional terms to the effective action which
break the continuous Galilean invariance of $S_{\rm sm}$ down to the
discrete space-group symmetries of the lattice.  This is accomplished
by including a {\sl periodic potential} for the bosons: 
\begin{equation}
  S_{\rm sm}
  \rightarrow \tilde{S}_{\rm rm} \equiv S_{\rm sm} + S_{\rm sm}^{\rm
    lattice},
\end{equation}
where
\begin{equation}
  S_{\rm sm}^{\rm lattice} = \int_{{\bf x}\tau} {\cal U}({\bf x})\,
  \vec{\phi}^* \cdot \vec{\phi}, 
\end{equation}
where ${\cal U}({\bf x})$ is an arbitrary periodic function with the
symmetries of the GSS lattice, i.e. ${\cal U}({\bf x}+{\bf a}_1) =
{\cal U}({\bf x}+{\bf a}_2) = {\cal U}({\bf x})$.  If desired, ${\cal
  U}({\bf x})$ could be specified by its Fourier coefficients at
reciprocal lattice vectors.  We will not, however, require a specific
form for the heuristic considerations of this section.

For the modified action $\tilde{S}_{\rm sm}$, a simple analysis
strongly suggests that the ``N{\'e}el'' state should occur if $V$ is
large and positive.  To see this, it is useful to rewrite
\begin{multline}
  U (\vec{\phi}\cdot\vec{\phi}^*)^2 + 
V(\vec{\phi}\cdot\vec{\phi})(\vec{\phi}^*\cdot\vec{\phi}^*) \\
=
  (U+V)(\vec{\phi}\cdot\vec{\phi}^*)^2 - V (i \vec{\phi}^* \times 
\vec{\phi})^2. \label{rewrite}
\end{multline}
Thus, positive $V$ indeed favors configurations with
$\vec{S}_1^{\rm tot} \propto i\vec{\phi}\times\vec{\phi}^* \neq 0$.  The
last term in Eq.~\ref{rewrite}\ can be decoupled using a
Hubbard-Stratonovich transformation,
\begin{equation}
  - V (i \vec{\phi}^* \times \vec{\phi})^2 \rightarrow -\vec{N} \cdot i
  \vec{\phi}^*\times\vec{\phi} + N^2/4V. \label{HS}
\end{equation}
We may envision a treatment in which the ``order parameter''
$\vec{N}$ is taken account in the saddle-point approximation, but the
bosons themselves are treated exactly.  The task then would be to find 
the ground state for the $\vec{\phi}$ bosons in the presence of an
arbitrary (presumed constant for simplicity) $\vec{N}$, and
subsequently to determine the optimal $\vec{N}$ by energy
minimization.  At the single-particle level, $\vec{N}$ appears
essentially as a ``Zeeman field'' splitting the triplet of bosons into 
three inequivalent states.  In the absence of the periodic potential,
these have single-particle energies
\begin{equation}
  \epsilon_m({\bf k}) = R + C k^2 - |\vec{N}| m,
\end{equation}
where $m=-1,0,1$ is the angular momentum projection of the boson along 
the $\hat{ N}$ axis (e.g. for $\vec{N} = |\vec{N}|\hat{ z}$, the three boson
eigenstates are 
$\phi_{\pm} = (\phi_x \pm i \phi_y)/\sqrt{2}$,
$\phi_0 = \phi_z$).  Thus with $|\vec{N}|\neq 0$, the $m=1$ state has lowest 
energy.  For $|\vec{N}|>R$, the ground state necessarily contains these
bosons.  If $|\vec{N}|$ is large, it is reasonable to ignore the $m=0,-1$
bosons which have larger energy and need not be present.  

In the absence of the periodic potential there is no alternative but
for these $m=1$ bosons to condense, giving rise to a state with either
$\vec{T}$ or $\Delta\vec{S}$ (or both) non-zero.  In the full
effective action, $\tilde{S}_{\rm sm}$, however, another possibility
exists.  Indeed, assuming $\phi_0=\phi_{-1}= 0$, $\tilde{S}_{\rm sm}$ 
describes a system of short-range {\sl interacting} bosons in a
periodic potential.  If the density of these bosons is sufficiently
large (i.e. the boson ``chemical potential'' $R-|\vec{N}|$ is sufficiently
negative), and ${\cal U}({\bf x})$ is strong, it may be energetically
favorable for one boson to ``localize'' in each minima of ${\cal
  U}({\bf x})$, more bosons being prevented from localizing by the
short-range repulsion $U$.  This is the desired bosonic Mott
insulator, which has the properties of the antiferromagnet, to wit, an
expectation value of the total spin on each dimer, without coincident
``Bose condensation'', i.e. $\langle\vec{T}\rangle = \langle
\Delta\vec{S}\rangle = 0$.  Clearly, the antiferromagnetic state requires
{\sl large} $|\vec{N}|$, and hence cannot be accessed by a continuous
transition from the dimer phase.  

\subsection{WISDW to AF transition}

Armed with the physical picture described above for the sub-model, we
now turn to a long-wavelength description of the analogous
Mott-insulating physics in the full GSS Hamiltonian.  In particular,
we focus on the question of the nature and order of a hypothetical
transition between the AF and WISDW states.  Because both phases
sustain N{\'e}el order, it is valid to assume from the onset an
expectation value $\vec{N}$ of the order parameter.  For simplicity,
we will neglect (quantum) fluctuations of $\vec{N}$.  By doing so, we
ignore the effects of the two antiferromagnetic magnon modes on the
critical properties.  We expect this to be a valid approximation for
several reasons.  First, both phases exhibit {\sl long-range}
spin-density-wave order, implying limited fluctuations of $\vec{N}$.
Second, the linear dispersion, $\omega \sim v_s |k|$, of
antiferromagnetic magnons implies that their characteristic
frequencies are much higher than those of the critical modes, which
scale approximately as $\omega \sim c |k|^2$.  Thus, on the long time
scales appropriate to the critical dynamics of the AF-WISDW
transition, the magnons are expected to be irrelevant.  An important
caveat is that, in the ordered (WISDW) phase, because the magnetic
wavevector becomes incommensurate, the magnon modes must become
involved.  We therefore expect that the coupling of magnons to the
critical modes is {\sl dangerously irrelevant}.  The treatment of this
section then suffices to understand the properties on the disordered
(AF) side of the transition and in the critical regime, but not at
very low $\omega,k$ on the ordered (WISDW) side.

The assumed constant $\vec{N}$ couples to the Bose fields as
\begin{equation}
  {\cal L}_{N-B} = -\vec{N}\cdot\left(i
  \vec{\phi}_1^* \times\vec{\phi}_1 - i\vec{\phi}_2^*\times \vec{\phi}_2 
\right),
\end{equation}
which, for $\vec{N} = |\vec{N}|\hat{ z}$ can be rewritten in the
angular-momentum basis as
\begin{equation}
{\cal L}_{N-B} = -|\vec{N}|\left( 
\phi_{1+}^* \phi_{1+^{\vphantom\dagger}} \! 
- \! \phi_{1-}^* \phi_{1-}^{\vphantom\dagger} \! 
- \! \phi_{2+}^* \phi_{2+}^{\vphantom\dagger} \! 
+ \! \phi_{2-}^* \phi_{2-}^{\vphantom\dagger} \right),
\end{equation}
with $\phi_{i\pm}$ defined as in the previous subsection.  Near the
AF-WISDW transition, the N{\'e}el order is well-developed, and a large
splitting exists between the lowest energy $\phi_{1+}$ and $\phi_{2-}$
modes.  It is therefore appropriate to integrate out the other four
``massive'' modes: $\phi_{10}, \phi_{1-}, \phi_{2+},\phi_{20}$.  The
remaining two fields $\phi_+ = \phi_{1+}$ and $\phi_- = \phi_{2-}$
constitute the order parameters for the quantum phase transition.

To proceed, we develop a Landau theory for this critical point using
the relevant symmetries.  The symmetries that remain unbroken in the
``disordered'' (AF) phase are: (1) the U(1) spin rotational symmetry
about the $\hat{ z}$ axis in spin space, (2) discrete spatial
reflection symmetries in the $x$ and $y$ directions, (3) a combined
spin-reflection ($\pi$ rotation around the $\hat{ x}$ or $\hat{ y}$
spin axis) with a simultaneous $\pi/2$ spatial rotation, and (4) a
combined $\pi/2$ spatial rotation and time-reversal operation.

Interestingly, the continuous U(1) spin-rotation symmetry acts as a
simple phase rotation of the order parameters,
\begin{equation}
\phi_\pm \rightarrow e^{\pm i\theta} \phi_\pm,
\end{equation}
where $\theta$ is an arbitrary U(1) phase.  Note that although there
are two complex order parameters, there is only a single U(1)
invariance. The symmetries (3) and (4) respectively correspond to the
transformation  
\begin{equation}
\begin{pmatrix}
x \\ y
\end{pmatrix}
\to
\begin{pmatrix}
y \\ -x
\end{pmatrix}
~ ;~
\begin{pmatrix}
\phi_{+} \\ \phi_{-}
\end{pmatrix}
\to
\begin{pmatrix}
\phi_{-} \\ -\phi_{+}
\end{pmatrix},
\end{equation}

and 
\begin{equation}
\tau \to -\tau ~;~
\begin{pmatrix}
x \\ y
\end{pmatrix}
\to
\begin{pmatrix}
y \\ -x
\end{pmatrix}
~ ;~
\begin{pmatrix}
\phi_{+} \\ \phi_{-}
\end{pmatrix}
\to
\begin{pmatrix}
-\phi_{-}^{*} \\ \phi_{+}^{*}
\end{pmatrix}.
\end{equation}

Taking into account all the symmetries, the general Landau form of the
effective Lagrange density, keeping constant terms and those leading
order in time and spatial derivatives, is
\begin{align}
&  {\cal L}_{AF-WISDW}  =  
\sum_{s=\pm} \phi_s^* 
\left[\partial_\tau -c(\partial_{x}^{2}+\partial_{y}^{2}) + \tilde{r}
\right] \phi_s 
\nonumber\\
&+\delta c
\left[\phi_{+}^{*}(\partial_{x}^{2}-\partial_{y}^{2})\phi_{+}
-\phi_{-}^{*}(\partial_{x}^{2}-\partial_{y}^{2})\phi_{-}\right]
\nonumber \\
& +d \left(\partial_{x}\phi_{+}\partial_{y}\phi_{-}+{\rm
    c.c.}\right)\nonumber\\ 
&  + \pi u\left( |\phi_+|^4 + |\phi_-|^4\right) + \pi v
  |\phi_+|^2|\phi_-|^2 
  + \pi w \left( \phi_+^2 \phi_-^2 + {\rm c.c.}\right), \label{AFWISDW}
\end{align}
where $\tilde{r}$ is proportional to the deviation from the
(mean-field) AF--WISDW critical point.  At the mean-field level, for 
$\tilde{r}>0$, bond-boson order is absent, and the system is in the AF
state, while for $\tilde{r}<0$, the bond-bosons are condensed, and the 
system is a WISDW.  Viewed as a field theory, 
the Lagrange density in Eq.~(\ref{AFWISDW}), like that of the original
DFT, describes complex Bose fields with dynamical critical exponent
$z=2$ at the Gaussian level.  Power counting thus implies the upper
critical dimension is again $d=2$, so that in three dimensions, the
AF-WISDW transition is mean-field-like and continuous.  In two
dimensions, such a transition may still occur, but fluctuation
corrections may again be significant.  Their effects are studied using 
the renormalization group in
Sec.~\ref{sec:AFWISDWRG}.

\section{Fluctuation Effects: Renormalization Group}
\label{sec:RG}

We have seen that, at the mean-field level, the dimer-boson approach
requires a minimum of two {\sl continuous} quantum phase transitions
to connect the dimer and AF states of a generic GSS model.  Moreover,
the explicit contruction of Landau theories appropriate to these
transitions demonstrates that the upper critical dimension for both
critical points is $d=2$.  Thus for the marginal case of two dimensions
fluctuation effects can be significant, and are fortunately amenable to
study using the renormalization group (RG).  In this section we
perform  RG analyses for both putative transitions.  In both cases,
because we are interested in the behavior precisely in the upper
critical dimension, we do not expect to find a non-trivial fixed
point (the analog of the Wilson-Fisher fixed point, should it exist,
merges with the Gaussian one as $d\rightarrow 2$ from below) .
Rather, fluctuation effects can either leave the Gaussian fixed point
stable, leading only to logarithmic corrections to mean-field critical 
behavior, or they may destabilize the Gaussian fixed point.  In the
latter case, the true behavior in the vicinity of the putative phase
transition is more subtle, and requires argumentation beyond the
simple RG.  

Due to its relative simplicity, we reverse the order of the previous
sections, and first discuss the AF-WISDW transition.  In this case, we
find that the Gaussian fixed point has a non-vanishing domain of
stability.  Thus the AF-WISDW critical point in $d=2$ displays, up to
logarithmic corrections, mean-field behavior: the correlation length
$\xi \sim \tilde{r}^{-\nu}$ with $\nu=1/2$, etc..  Next, we turn to
the study of the transition between the Dimer phase and the WISDW, and
more generally to the other mean-field like phase described in section
\ref{sec:smmft}.  The conclusion of exceptionally involved
calculations is in stark contrast to that for the AF-WISDW transition.
Regardless of the ``bare'' values of the coupling constants, without
excessive fine-tuning, fluctuations always drive the system away from
the Gaussian fixed point.  Thus mean-field behavior definitely does
not apply in the naive critical regime, and moreover a more complex
critical scenario may obtain.  We argue that a natural 
possibility is that fluctuations nucleate an intermediate phase
between the dimer and WISDW states.  This intermediate phase is very
likely to be an exotic, ``fractionalized'' state without long-range
magnetic order but with elementary excitations carrying spin-$1/2$, as 
suggested recently by a completely different large-$N$
approach.\cite{subir-private}\

\subsection{RG for the AF--WISDW transition}
\label{sec:AFWISDWRG}

The AF-WISDW transition is described by the Lagrange density in
Eq.~\ref{AFWISDW}.  To determine the modifications to mean-field
behavior, it is sufficient to carry out an RG analysis to one-loop in
the quartic interactions $u$, $v$, and $w$.  The classical RG
techniques needed to derive the equations can be found in numerous
textbooks\cite{Zinn}.  A convenient way to proceed
is the so-called floating cut-off procedure.  For simplicity, we will
consider in this discussion a hard cut-off, where the fields
$\phi_\pm$ are defined for momenta $0<|{\bf q}|<\Lambda $, although in
the following we will switch to other regularizations.  Let us
consider the partition function $Z=\int
d[\phi_\pm]d[\phi_\pm^{*}]e^{-S(\phi_\pm,\phi_\pm^{*})}$.  Under a
change of cut-off $\Lambda \to \Lambda'=\Lambda/b$, we can split the
measure of integration $d[\phi_\pm]$ into an integration over 'fast
modes' $\phi^{>}_\pm$ and 'slow modes' $\phi^{<}_\pm$, where
$\phi^{>}_\pm({\bf q})$ is non zero only for $\Lambda/b<|{\bf
  q}|<\Lambda $, and $\phi^{<}_\pm({\bf q})$ for $|{\bf
  q}|<\Lambda'=\Lambda/b$.  The averaging over the 'fast modes'
defines the remaining interactions between the slow modes with {\it
  renormalized} couplings, once we have rescaled the fields and momenta
according to
\begin{equation}
\phi_{i}^{<}\to b^{\zeta}\phi_{i}' 
\quad ;\quad 
x\to b x'
\quad ;\quad 
t\to b^{z} t' .\label{eq:rescaling}
\end{equation}
The fast mode integration is performed using the cumulant expansion,
which to second order reads
\begin{equation}\label{cumulantshort}
\delta S=
\langle S_{I}\rangle^{>} -\frac{1}{2!}\langle S_{I}^{2}\rangle^{>}_{c},
\end{equation}
where $S_{I}$ is the non-quadratic part of the action in
Eq.~\ref{AFWISDW}, and $\langle \rangle^{>}_{c}$ corresponds to the
connected averaged over the 'fast modes' using the weight defined by
the quadratic Lagrangian (i.e. setting $u=v=w=0$).  We do this for
infinitesimal rescaling $b=e^{dl}$, where $dl \rightarrow 0^+$ is a
positive infinitesimal, and $l$ is the cumulative
(logarithmic) rescaling. 

The freedom of choosing $\zeta$ and $z$ during rescaling in
Eqs.~\ref{eq:rescaling}\ is sufficient to keep the {\sl isotropic}
part of the quadratic Lagrangian fixed as the fast modes are
integrated out, and we choose this convention.  The Lagrangian has,
however, two additional parameters, $\delta c$ and $d$, which
parameterize deviations from spatial isotropy (rotational invariance).
With some effort, the one-loop RG for the quartic interactions can be
done to {\sl all orders} in $\delta c$ and $d$, since these couplings
are quadratic in the fields.  However, the freedom to rescale under
the RG is insufficient to maintain scale-independent values of $\delta
c$ and $d$ at the fixed point, so these couplings will themselves obey
flow equations.  From prior experience, we expect $\delta c$ and $d$
to in fact renormalize to {\sl zero}, so that ultimately they may also
be treated perturbatively.

\begin{figure}[htb]
\centerline{\fig{6cm}{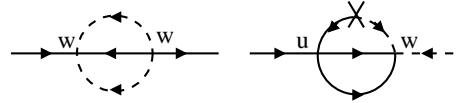}}
\caption{Leading-order two-loop contributions to $\delta c$ and $d$.}
\label{fig:twoloop}
\end{figure}

To verify this, we consider the leading-order renormalization of
$\delta c$ and $d$.  This occurs first at {\sl two loop} order, by the
diagrams in Fig.~\ref{fig:twoloop}.  Taking into account their effects 
on both the isotropic terms and $\delta c$ and $d$, we obtain the flow 
equations
\begin{subequations}
  \begin{align}
    \partial_{l} (\delta c) &= -\frac{4}{27}w^{2}(\delta c), \\
    \partial_{l} d &= -\frac{1}{9}\left[w^2 +\frac{1}{8}uw\right]d.
  \end{align}
\end{subequations}
Hence for $\delta c$ always flows to zero, as does $d$ in the
non-vanishing basin of attraction $uw>-8w^2$.  Thus the isotropic
point is locally (marginally) stable and we can restrict our study to
an isotropic propagator with $\delta c=d=0$.

The one-loop RG equations for
the quartic coupling constants are then readily obtained and read 
\begin{subequations}\label{RG-WISDW}
\begin{align}
\partial_{l} u &= -u^{2}-w^{2},\\
\partial_{l} v &= -v^{2}-4w^{2},\\
\partial_{l} w &= -(u+v) w.
\end{align}
\end{subequations}
Hence, while $u$ and $v$  always renormalize downwards to
negative values, $w$ can either decrease to zero or diverge 
depending on the sign of $u+v$. 
%
A schematic RG flow of the above
equations (\ref{RG-WISDW}) is shown on figure \ref{fig:RG-WISDW}. 
\begin{figure}
\psfrag{w}{$w$}\psfrag{u+v}{$u+v$}
\centerline{\fig{8cm}{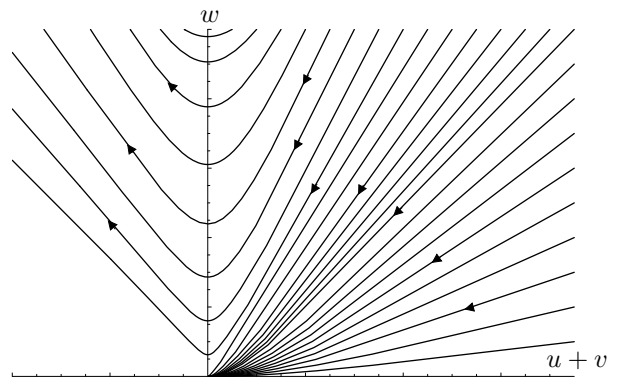}}
\caption{RG flow for $v_{0}=.1,w_{0}=.1$
and $-0.1<u_{0}<0.65$. The flow is shown in
coordinates $u+v,w$.}
\label{fig:RG-WISDW}
\end{figure}
As seen in the figure, there is a finite region of bare values for
$u,v,w$ for which the RG flows lead back to the Gaussian fixed point.
Analytically, one can show that asymptotically stable solutions exist
for which $u+v \simeq 2/l$, $u-v\simeq u_0 /l^2$, and $w \simeq
w_0/l^2$, where $u_0$ and $w_0$ are constants.  The separatrix
surface between the two phases, shown on the figure, can be shown to
correspond to the asymptotic behaviour around the origin : $u+v\simeq
l^{-1}[1+\frac{1}{2}(\ln l)^{-1}+{\mathcal O}(\ln^{-2} l)]$, $u-v\simeq
l^{-1}[1-\frac{5}{6}(\ln l)^{-1}+{\mathcal O}(\ln^{-2} l)]$, $w\simeq
\pm l^{-1} [\frac{1}{\sqrt{6}}(\ln l)^{-1}+{\mathcal O}(\ln^{-2} l)]$.

The renormalization study thus confirms the existence of a continuous
transition between the AF and WISDW phases, which is, up to
logarithmic corrections, of the mean-field type.  As remarked above,
we expect the antiferromagnetic magnons to be (dangerously) irrelevant
at the critical point we found, but this expectation remains to be
confirmed by detailed calculations beyond the scope of this paper.

\subsection{RG for the dimer--WISDW transition}

We now turn to the effect of quantum fluctuations on the dimer-WISDW
transition, described by the action (\ref{def-s}).  This requires
extending the RG analysis of the previous subsection to the much more
formidable Lagrangian fo the DFT.  The crucial and new
feature of the DFT (\ref{def-s}) relative to the reduced Lagrangian
for the AF-WISDW transition, is obviously the cubic terms
(\ref{s3}) which are directly related to the geometry of the SS
lattice and the exactness of the SS dimer state.   Much of the technical
difficulties of the RG approach can be related to the presence of this
term, which as we saw in the previous sections, will favor spatially
non-uniform and non-collinear fluctuations.

On the technical side, the huge number of coupling constants in the
action (\ref{def-s}) can be conveniently handled using tensorial
notation.  These notations, the technical details of our approach, and
the explicit RG equations are postponed to the appendix
\ref{sec:appendixRG}.  We here focus on the spirit of the method and
the structure of the results. These are used to numerically study the
scaling behaviour of the action (\ref{def-s}) perturbatively in the
couplings of the cubic and quartic terms.  We find that these
interaction coefficients always grow in magnitude under the RG,
indicating a behaviour different from the mean-field one.  The
technical aspects of these calculations are somewhat formidable, and
the less strong-willed reader may wish to skip at this point to the
end of this subsection, where we discuss the {\sl results and
  consequences} of the RG calculations.  Others should press on for a
brief summary of the calculations, and true afficionados, if any there 
be, will find further details in appendix~\ref{sec:appendixRG}.

\subsubsection{Idea of the method}

As for the AF-WISDW transition, we simplify the calculations by
working perturbatively in the violations of anisotropy $\delta c$ and
$d$.  In this way, one may derive one-loop RG equations for the
tensorial couplings $G,U,V,W$.   These are represented graphically in 
Fig.~\ref{fig:diagrams-S}.
\begin{figure}
\psfrag{k1}{}\psfrag{k2}{}\psfrag{k3}{}\psfrag{k4}{}
\centerline{
\diag{6mm}{1.6cm}{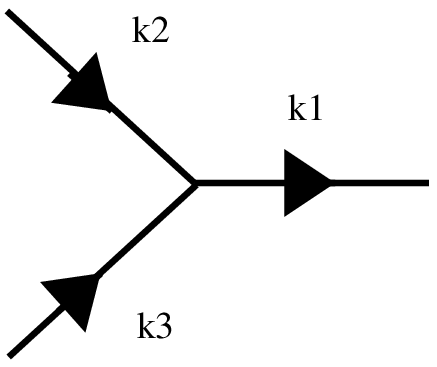} : $G({\bf q}_{1},{\bf q}_{2})$
\diag{7mm}{2.5cm}{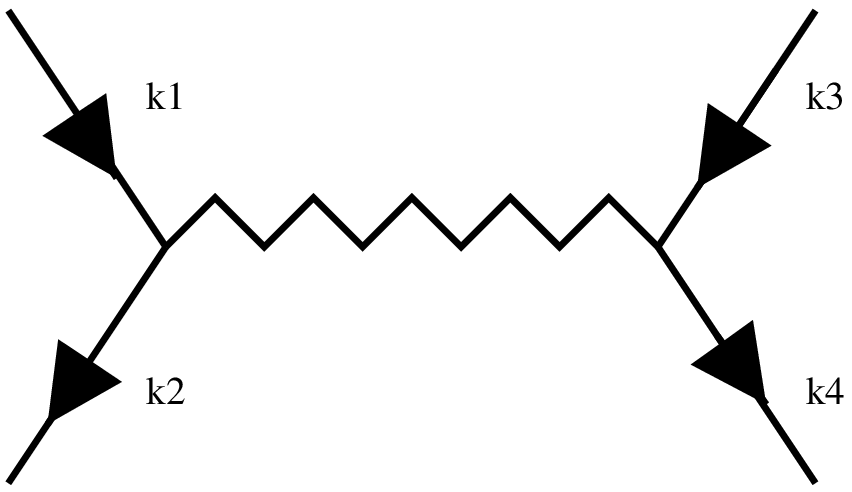} : $U$}
\centerline{
\diag{7mm}{2.5cm}{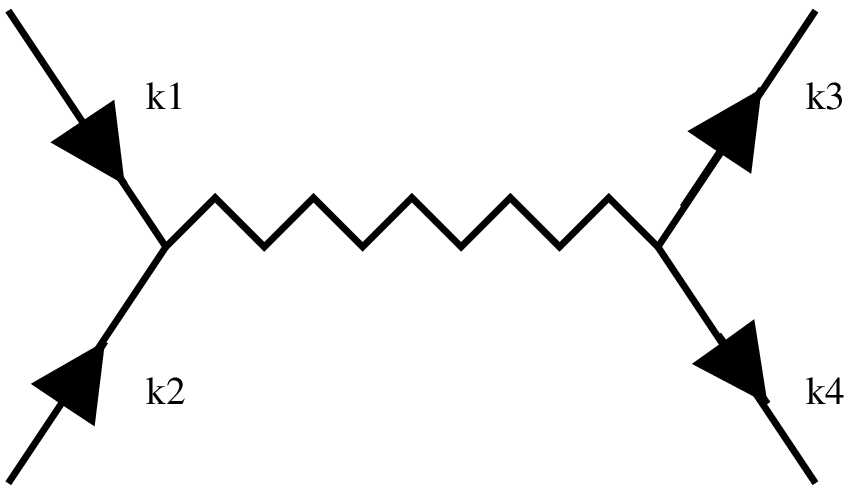} : $V$ \hspace{.4cm}
\diag{7mm}{2.5cm}{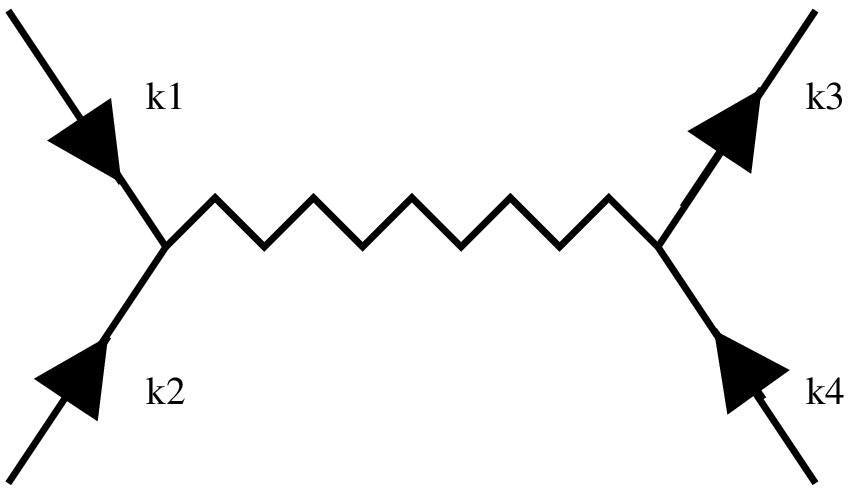} : $W$}
\caption{Diagrammatic representation of terms in $S^{(3)}$ and
$S^{(4)}$. Ingoing arrows correspond to {\it e.g} $\vec{\phi}^{*}$. 
Complex conjugates of the above diagrams have been omitted for clarity. }
\label{fig:diagrams-S}
\end{figure} 

The specificity of the field theory (\ref{def-s}) is obviously the
cubis term (\ref{s3}). It induces most of the differences between
our results and the more conventional renormalization of a $\vec{\phi}^{4}$
field theory.  As the first of its consequences, we must consider
up to the fourth term in the cumulant expansion that defines the
renormalised action :
\begin{equation}\label{cumulant}
\delta S=
\langle S_{I}\rangle^{>} -\frac{1}{2!}\langle S_{I}^{2}\rangle^{>}_{c}
+\frac{1}{3!} \langle S_{I}^{3} \rangle^{>}_{c} 
-\frac{1}{4!}\langle S_{I}^{4} \rangle^{>}_{c}, 
\end{equation}
where $S_{I}=S^{(3)}+S^{(4)}$.  In each of these terms, only one-loop
diagrams are considered. The result can be organized by collecting the
contribution to the renormalized propagator and $G,U,V,W$ (shown as
black boxes on figure \ref{fig:diag-cumulant}).  The different 1-loop
diagrams are shown in figure \ref{fig:diag-cumulant}.

\startlargeeq 

\vspace{-1cm}
\begin{figure}[!t]
\psfrag{k1}{}\psfrag{k2}{}\psfrag{k3}{}\psfrag{k4}{}\psfrag{k5}{}\psfrag{k6}{}
\psfrag{A}{}\psfrag{B}{}\psfrag{C}{}\psfrag{D}{}\psfrag{E}{}\psfrag{F}{}
\psfrag{p}{}\psfrag{q}{}\psfrag{m}{}\psfrag{n}{}
\begin{align*}
\diag{3mm}{2.5cm}{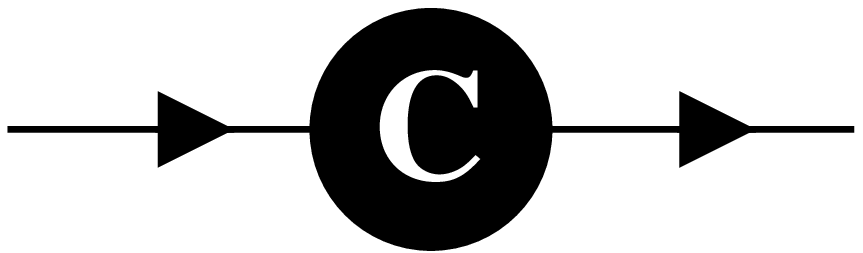}=&
\diag{0mm}{2cm}{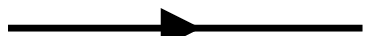} 
-2 \diag{5mm}{2.5cm}{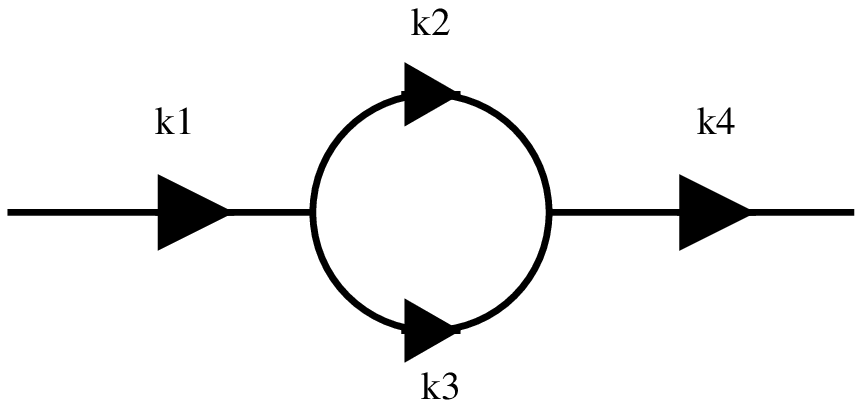}
\\
\diag{7mm}{1.8cm}{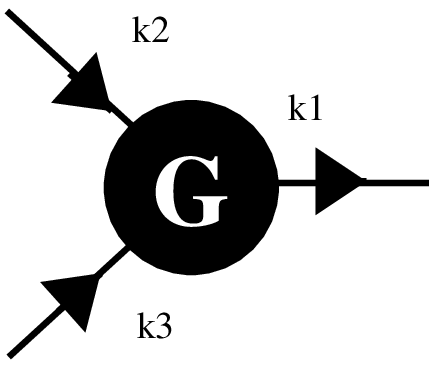}
 =&
\diag{7mm}{1.8cm}{gtensor}
-2~G U
\diag{8mm}{2.2cm}{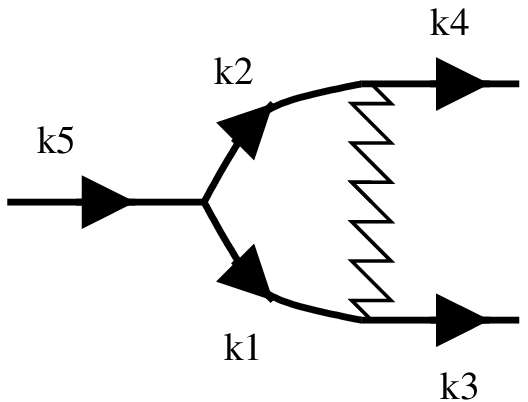}
-4~ G W
\diag{8mm}{2.2cm}{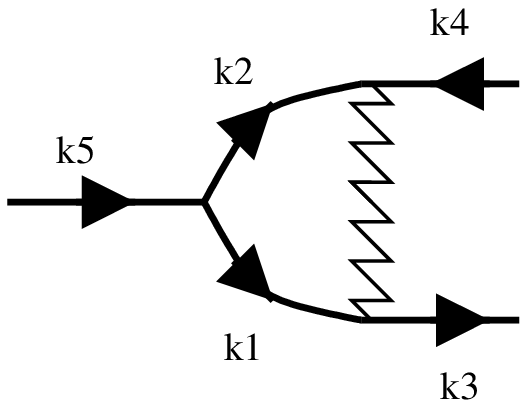}
+8~G^{3}
\diag{8mm}{2cm}{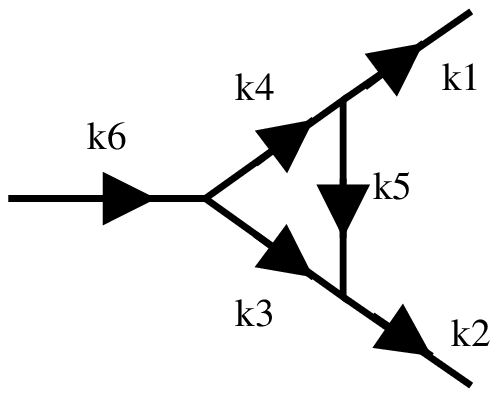}
\\
\diag{6mm}{2cm}{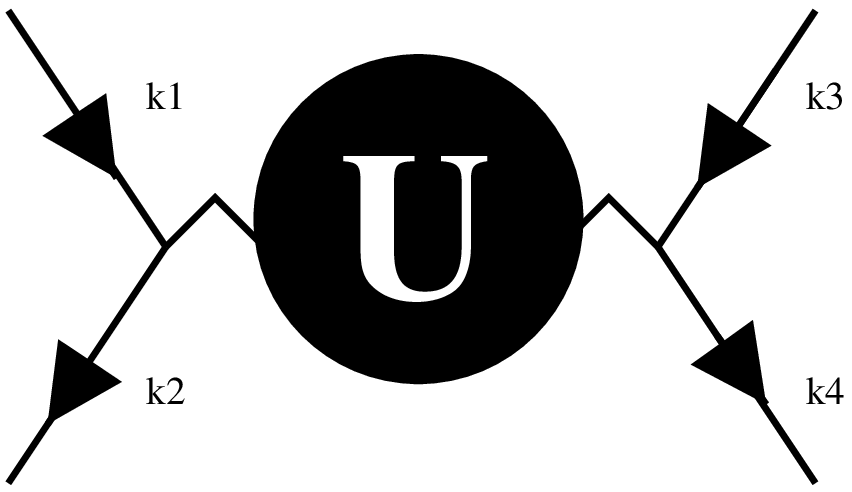} =&
\diag{6mm}{2cm}{U-tensor}
-2 U^{2}\diag{7mm}{1.5cm}{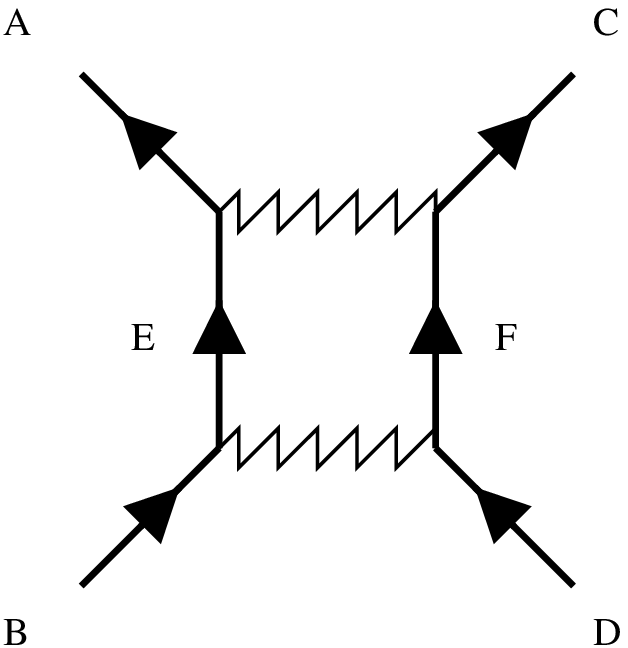}
-4W^{2}\left(
 \diag{5mm}{2.5cm}{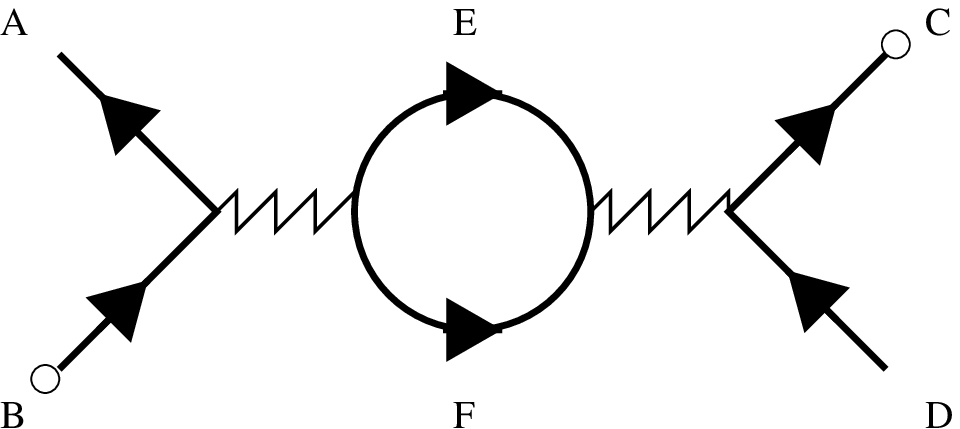}
+ \diag{7mm}{1.5cm}{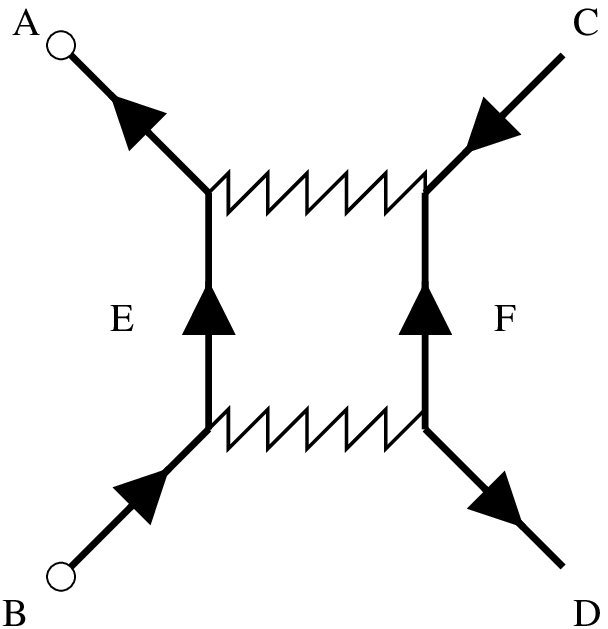}
 \right)
-16~G^{4} 
\diag{7mm}{1.5cm}{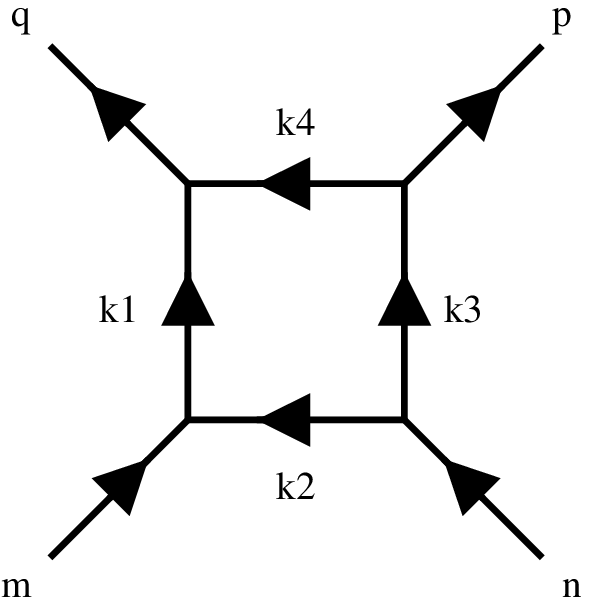}\\
\diag{6mm}{2cm}{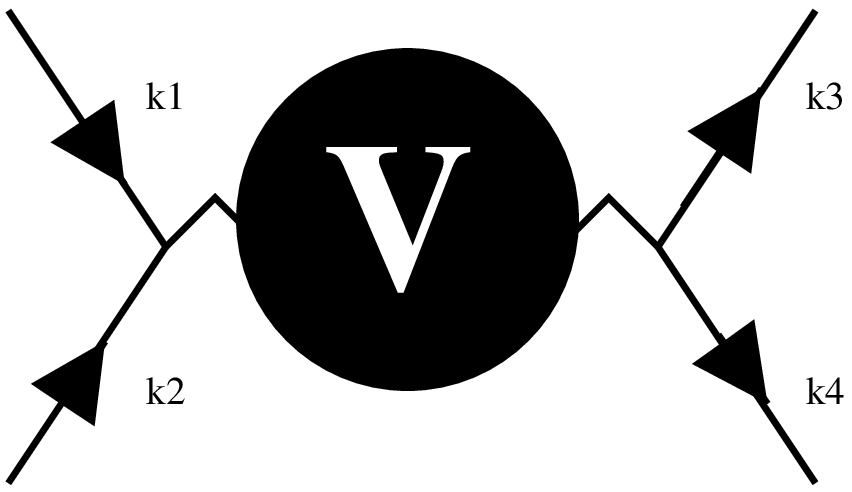} =&
\diag{6mm}{2cm}{V-tensor}
-4~V^{2}\diag{5mm}{2.5cm}{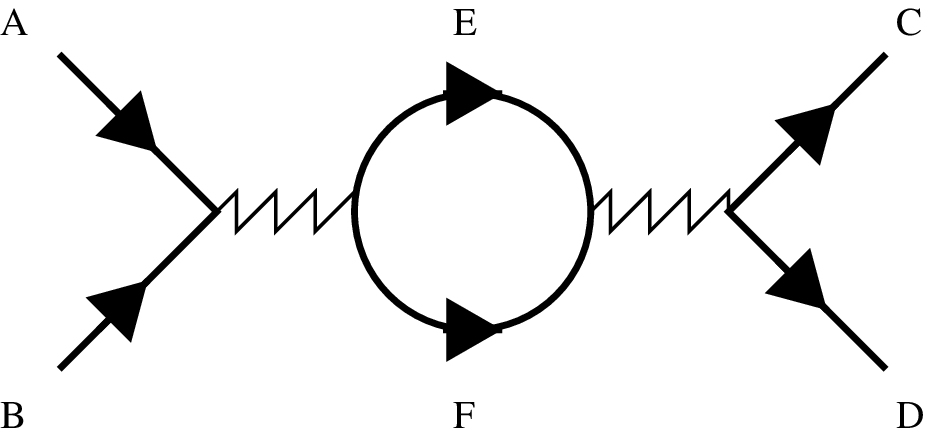}
-4W^{2} \diag{7mm}{1.5cm}{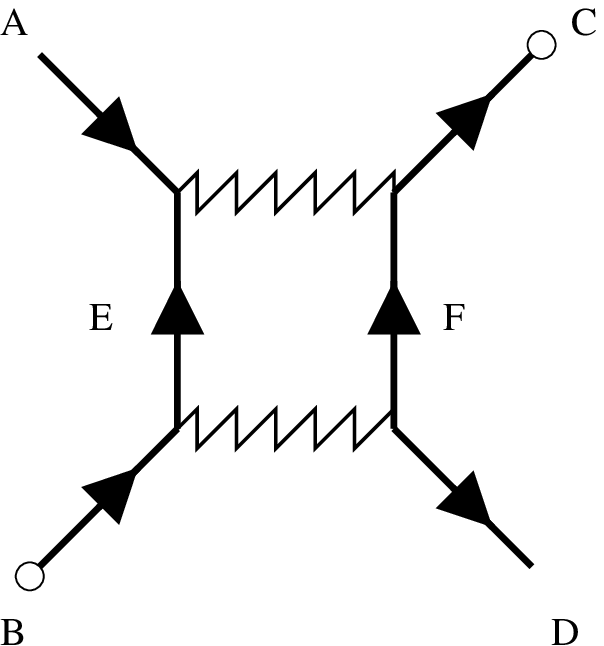}
-4 UV \diag{5mm}{2.2cm}{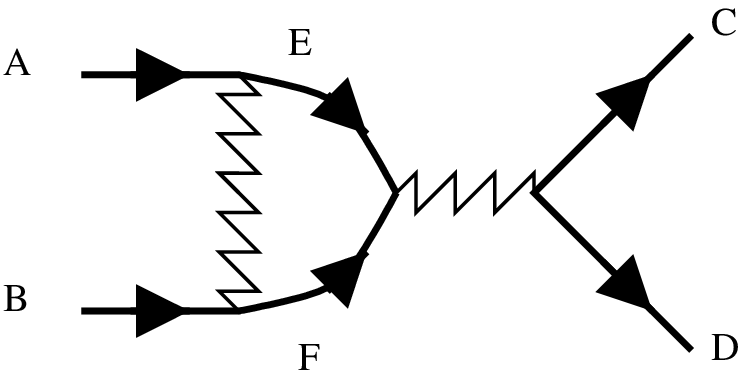}
-16~G^{4} 
\diag{7mm}{1.5cm}{GGGG-2}
\\
\diag{6mm}{2cm}{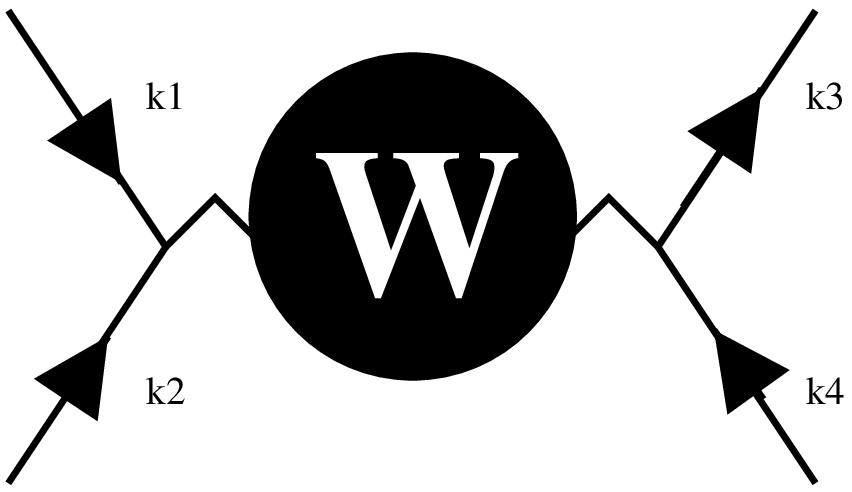} =&
\diag{6mm}{2cm}{W-tensor}
- 2UW~\diag{5mm}{2.2cm}{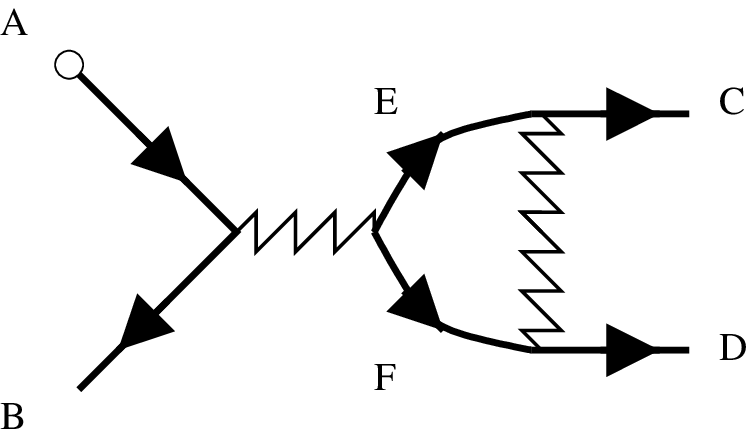}
-VW\left(
 2~\diag{5mm}{2.5cm}{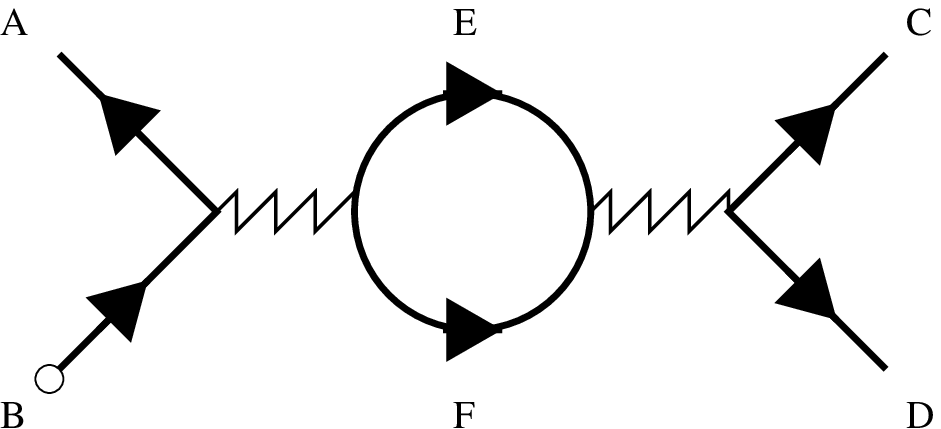}
+4~\diag{5mm}{2.2cm}{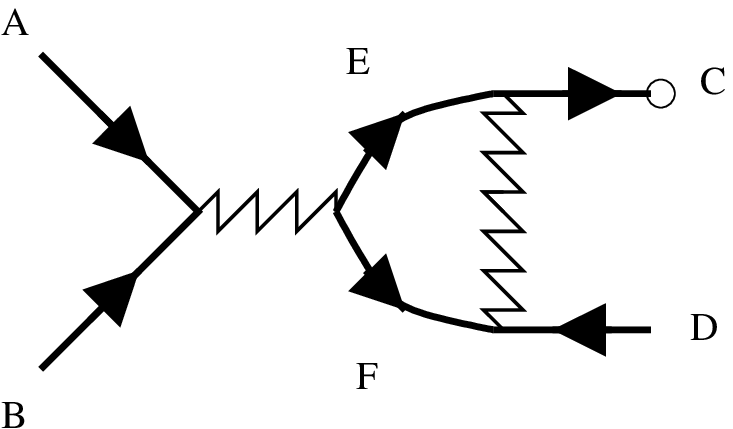}+cc
\right)\\
&
-4 UW  \diag{7mm}{1.5cm}{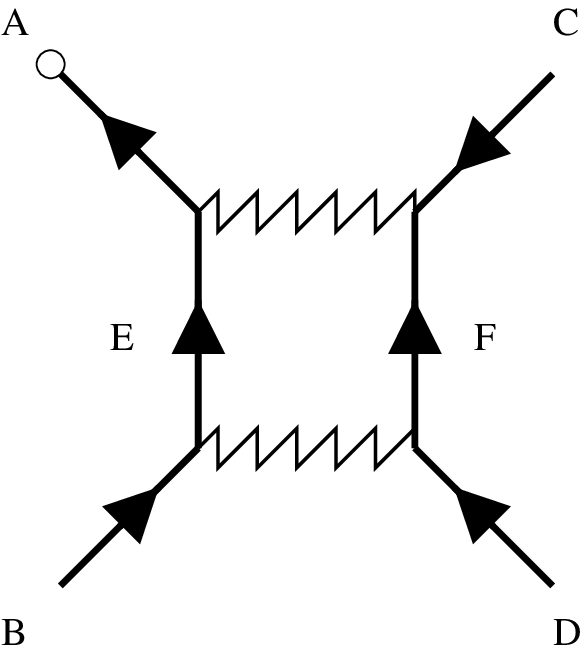}
-16~G^{4}
\diag{7mm}{1.5cm}{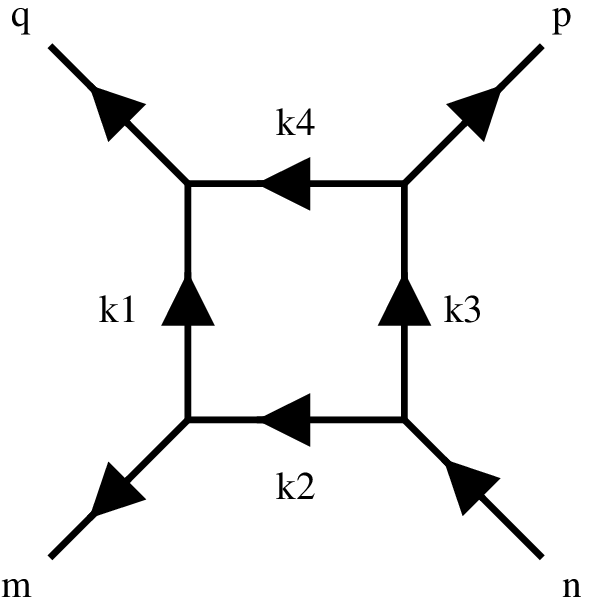}
\end{align*}
\caption{Diagramatic representation of the contributions of order one
loop to the renormalised coupling tensors $G,U,V,W$ and the propagator
$C$.}
\label{fig:diag-cumulant}
\end{figure}
\stoplargeeq 

One can notice that the peculiar cubic term '$G$' gives rise to
contributions both to the anomalous dimensions and {\it all} the
coupling constants. In particular, it renormalizes itself to order
$G^{3}$ and the quartic couplings to order $G^{4}$.  
The unusual derivative nature of these
interactions requires consideration of non-trivial vertices at
non-vanishing external momentum even in the one-loop RG, which is
extremely unusual.

Due to their complexity, the different contributions are presented
in the appendix  \ref{sec:appendixRG}. 
We here focus on the structure of the RG equations. 
{}From the result of appendix \ref{sec:appendixRG}, we find the value
of $\zeta$ and the dynamical exponent $z$ to order one loop :  

\begin{eqnarray}
\zeta &=&
-1 -\frac{1}{8\pi }\left(g_{0}^{2}+g_{2}^{2} \right)
-\frac{1}{16\pi}\left(g_{1}+g_{4}\right)^{2}\\
\nonumber 
&&+\frac{1}{16\pi }\left[ \delta c(g_0^2-g_2^2)+d(g_0+g_2)(g_1+g_4)\right],
\\
z &=& 
2+\frac{1}{8\pi}
\left(g_{0}^{2}+ g_{2}^{2}+g_{1}^{2}+g_{4}^{2}\right)\\
\nonumber 
&&-\frac{1}{16\pi}\left[
\frac{1}{2}\delta c(g_{0}^{2}-g_{2}^{2}+g_{4}^{2}-g_{1}^{2})
+d(g_{0}g_{4}+g_{1}g_{2}) \right]. 
\end{eqnarray}

 As shown in the appendix, using these exponents and the direct
contributions of the diagram to the couplings $G,U,V,W$, we obtain the
non-linear RG equations describing the scaling behaviour of the
couplings of the field theory (\ref{def-s}). 
For simplicity, we present here only the structure of these
equations. The brave and bold reader who might be interested in the
detailed definition of the tensorial coefficients of these equations
can find the explicit equations at the end of the appendix.  

\subsubsection{RG results and analysis}

The one-loop RG equations are
\begin{subequations}\label{RGeqs}
\begin{align}
&\partial_{l} g_{i} =
 -\frac{1}{32\pi} \Gamma^{(g,i)}_{jkl}~g_{j}g_{k}g_{l}
  - \frac{1}{4\pi}\Theta^{(i)}_{jk}~g_{j}u_{k},\\
&\partial_{l}u_{i}=
\left[-\frac{3}{8\pi }\left(g_{0}^{2}+g_{2}^{2} \right)
-\frac{1}{8\pi}\left(g_{1}+g_{4}\right)^{2}
 \right]u_{i}\\
&
-\frac{1}{64 \pi}\Gamma^{(u,i)}_{jklm}~g_{j}g_{k}g_{l}g_{m}
-\frac{1}{2\pi}
\left(\Lambda^{(1,i)}_{jk}u_{j}u_{k}
     +\Lambda^{(2,i)}_{jk}w_{j}w_{k} \right),  
\nonumber \\
&\partial_{l}v_{i}=
\left[-\frac{3}{8\pi }\left(g_{0}^{2}+g_{2}^{2} \right)
-\frac{1}{8\pi}\left(g_{1}+g_{4}\right)^{2}
 \right]v_{i}\\
&\quad 
-\frac{1}{2\pi}
\left(\Lambda^{(3,i)}_{jk}u_{j}v_{k}
     +\Lambda^{(4,i)}_{jk}v_{j}v_{k}
     +\Lambda^{(5,i)}_{jk}w_{j}w_{k} \right) \nonumber \\
&\quad 
-\frac{1}{64 \pi}\Gamma^{(v,i)}_{jklm}~g_{j}g_{k}g_{l}g_{m}, \nonumber \\
&\partial_{l}w_{i}=
\left[-\frac{3}{8\pi }\left(g_{0}^{2}+g_{2}^{2} \right)
-\frac{1}{8\pi}\left(g_{1}+g_{4}\right)^{2}
 \right]w_{i}\\\nonumber 
&
-\frac{1}{64 \pi}\Gamma^{(w,i)}_{jklm}~g_{j}g_{k}g_{l}g_{m}
-\frac{1}{2\pi}
\left(\Lambda^{(6,i)}_{jk}u_{j}w_{k}
     +\Lambda^{(7,i)}_{jk}v_{j}w_{k} \right).  
\end{align}
\end{subequations}
 
As expected, since $d=2$ corresponds to the upper critical dimension
of the DFT (\ref{def-s}), the linear terms vanish in the right hand
side of the above equations. All the perturbing operators of the free
action thus correspond to either marginally relevant or marginally
irrelevant perturbations. Due to the complexity of these non-linear
flow equations, it is extremely difficult to determine the stability
of the Gaussian fixed point analytically.  Instead, we have
concentrated on a numerical integration of Eqs.~(\ref{RGeqs}),
supplemented with analytical treatment in particular limits.  After a
concerted effort to locate a stable basin of attraction, we have
concluded that the ultimate fate of the RG, is a ``run-away'' flow in
which several couplings tend to large (absolute) values as $l
\rightarrow \infty$.  We believe this to be true for generic bare
parameters, i.e. for all initial conditions except for various special
sets of parameters of measure zero in the space of marginal coupling
constants.

The driving force for this instability is most easily illustrated for the
simplified RG flows which apply to the sub-model, Eqs.~\ref{RG-submodel}\ of
appendix \ref{sec:appendixRG}.  The most significant aspect of these
equations is the RG 
flow for the cubic interaction,
\begin{equation}
  \partial_l G = \frac{3}{16\pi }G^3,
\end{equation}
which is {\sl independent} of the other coupling constants.  This
equation immediately indicates the flow of $G$ to large values,
irrespective of its sign.  As it does so, it tends to drive $U$ and
$V$ negative, leading to their divergence as well: in fact, it can be
shown from Eqs.~\ref{RG-submodel}\ that $U$ and $V$ always diverge
{\sl before} $G$. 
Moreover, this instability is not generic in the absence of the cubic
interaction: for $G=0$ there is  a finite basin of
attraction for the Gaussian fixed point in the space of $U,V,W$.

Although the decoupling of the flow of the cubic 
interactions is not shared by the RG flows in the full DFT, the
remaining properties of the sub-model RG are indeed common to the DFT.
In particular, if all cubic interactions are {\sl fine-tuned} to zero, 
there is a stable basin for the Gaussian fixed point.  In a generic
situation, however, with non-vanishing cubic terms, the flow is always 
to strong coupling.  If the bare cubic couplings are taken initially
small, the RG flows for the quartic interactions initially tend
towards zero, but the cubic couplings slowly increase as this
proceeds.  Eventually, the feedback of the cubic terms into the RG for 
the quartic couplings drives one or more of these interactions into an 
obviously unstable regime, and the quartic interactions tend to large
values.  Like in the sub-model, it appears that the quartic couplings
diverge before the cubic ones.  

While strictly speaking such run-away RG flows indicate only the
failure of the perturbative RG analysis, we can get some physical
insight into the failure of the RG by considering the nature of the
diverging RG.  It is clear from the above discussion that the
divergence is driven by the cubic coupling constants.  Because each
involves a triple product, such terms contribute to the action {\sl
  only} for non-collinear configurations of the fields.  Their role in
the perturbative instability thus indicates that non-collinear
magnetic correlations (or at least fluctuations) are significant in
this critical regime.  If the RG flows are followed out to the
neighborhood of their actual divergence (this is justified for small
bare couplings -- see the discussion in Refs.~\onlinecite{Lin,etc}),
however, the quartic interactions become much larger than the cubic
ones.  Because the quartic terms do not involve spatial derivatives,
this hints that the incipient ordering may in fact be {\sl
  commensurate} (see, e.g. Sec.~IIIC).  Finally, the flow to negative
quartic couplings suggests that the dominant field configurations
(with largest effective action) have a non-zero amplitude in the
putative critical regime.

Based on these arguments, it is natural to suspect that the behavior
in this region of the phase diagram may be modeled a {\sl Non-Linear
  $\sigma$ Model} (NL$\sigma$M) , in which the order-parameter
amplitude is everywhere (and when) non-vanishing and hence
approximated as fixed.  The {\sl orientation} of the order parameter
may, however, fluctuate.  To determine the appropriate NL$\sigma$M, we
must specify the order parameter space.  In order to
incorporate the tendency towards antiferromagnetism favored by the $J_2$
spin-spin interaction, a natural choice is to consider configurations
of $\vec{\phi}_{1,2}$ which sustain local N{\'e}el order.  As explained in 
Sec.~IIIC, such states have order parameters of the form
\begin{equation}
  \vec{\phi}_1 = \vec{\phi}_2^*= A\hat{ e}_1 +
iB\hat{ e}_2, 
\end{equation}
where in general $A$ and $B$ are fixed constants, while $\hat{ e}_1$
and $\hat{ e}_2$ are arbitrary {\sl orthogonal} unit vectors,
$\hat{ e}_1\cdot\hat{ e}_2=0$, $\hat{ e}_1^2=\hat{ e}_2^2=1$.  The
NL$\sigma$M space is defined by the set of all such orthogonal
vectors, constraining the effective action by the spin-rotational and
lattice invariances of the GSS model.  Closely related NL$\sigma$Ms have
been derived and studied previously for non-collinear magnets in
Refs.~\onlinecite{Subir}.  In addition to ordered phases, they are
believed to sustain a magnetically disordered phase with {\sl
  fractional} (i.e.spin-$1/2$) excitations, ``spinons''.  We thus
guess that the flow of the RG to strong coupling may be indicative of
such a fractionalized phase intervening between the WISDW and dimer
states.  Because the strong-coupling RG flow is, of course, not truly
controlled, alternative scenarios are possible, including the less
interesting possibility of a direct first-order transition between the 
latter two phases.

\section{Summary}
\label{sec:summary}

In this paper we have demonstrated the utility of the dimer-boson
representation for quantum spin models.  The approach enables a simple 
derivation of critical field theories, particularly suitable for
systems with explicit dimerization.  The coarse-grained dimer-boson
order parameters describe a wide range of magnetic phases with complex 
local orders.  In frustrated magnetic models such as the GSS models,
this description gives a natural means to understand a diverse set of
spin orderings that naturally obtain.  

For the particular class of GSS models on which we have focused, the
method indicates the absence of a direct dimer-AF quantum critical
point, but leads to a two-stage transition through the intermediate
WISDW phase.  With three-dimensional coupling, the two transitions
from this phase may be continuous and mean-field-like, while in
strictly two dimensions quantum fluctuations destroy mean-field
behavior near the putative dimer-WISDW critical point.  We have argued
on physical grounds that the likely alternative is the existence of a 
``fractionalized quantum paramagnet'' between the dimer and WISDW
states.  We cannot rule out however other improbable scenarios. In
particular, our 1-loop RG study was based on some technical assumption
of 'weakly anisotropic' bosonic propagator which may be invalidated in
some regimes.   

More generally, the dimer-boson theory explored here has a very
interesting field-theoretic structure, and may lead to a variety of
potential extensions in the future.  One might wish to derive a more
precise complete MFT study of the full action (\ref{def-s}).  The
dimer boson representation could likely be fruitfully applied to other
frustrated magnetic models.  We leave these and other extensions to
future works.
 
{\it Acknowledgements}
\par

Thanks to R. Singh for stimulating our interest in this problem so
long ago, and to M. P. A. Fisher for discussions.  This research was
supported by the NSF CAREER program under Grant NSF-DMR-9985255, 
by the Sloan and Packard Foundations, and by National Science
Foundation under Grant No. PHY99-07949.

\appendix

\section{Explicit RG equations}
\label{sec:appendixRG}

In this appendix, we present briefly the excruciating  derivation
of the scaling equations for the model (\ref{def-s}) and the sub-model
(\ref{eq:sm}). 
The first step in the derivation of the RG equations 
is to determine all the necessary 1-loop terms in the
cumulant expansion (\ref{cumulant}).  This is conveniently achieved
graphically using the conventions of Fig.~\ref{fig:diagrams-S},
associated with the tensorial notations defined below. 
The result is shown in Fig.~\ref{fig:diag-cumulant}\ as the
definition of formal renormalized couplings and propagator
(represented by black boxes). In these
diagrams, each 'internal line' corresponds to a contraction between two
'fast modes', the external legs corresponding to the remaining slow
modes. 

The next step in this approach is to explicitly average over the
internal fast modes for each of these diagrams.  Special care has to
be given here to the diagrams involving a cubic vertex (from
Eq.~(\ref{s3})) due to their explicit and anisotropic momentum
dependance. The choice of the cut-off can then be of some importance.
In principle, any universal feature (like ratios between the
cofficients in the RG equations) is independent of the precise choice
of {\sl smooth} cut-off we choose.  However in practice it is often
useful to use a sharp cut-off, and one must to be careful not to spoil
the result by such a choice.  The first and natural choice is a sharp
isotropic cut-off function in momentum space : the fast modes then
correspond to $\Lambda e^{-dl}<|{\bf q}|<\Lambda$. We then have to
consider the dominant terms for large $\Lambda$ in the integrals
corresponding to the diagrams of fig. \ref{fig:diag-cumulant}.
Instead we will follow a more 'field-theoretical' route.   We have
checked term by term that these two methods give exactly the same
result, which reinforces our confidence in the result.

In this approach, the regularized propagator $C^{-1}_{0}({\bf
  q},\omega)$ is expressed as
\begin{equation}\label{cut-off}
\int_{\Lambda^{-2}}^{+\infty} dt~ e^{-t (cq^{2}+iZ\omega )}. 
\end{equation}
By differentiang the end result of each diagram with respect to
$\partial /\partial \ln \Lambda$, we obtain the desired result.     

Following this method, the contributions of the diagrams from
Fig.~\ref{fig:diag-cumulant}\ can be readily obtained for the sub-model
Eq.~(\ref{eq:sm}) for which the meaning of the conventions
(Fig.~\ref{fig:diagrams-S}) is obvious. We will thus first present the
results for this sub-model before turning back to the full DFT and the
associated tensorial notations.  For completeness, we will also
describe in detail the contribution from the most specific diagram of
Fig.~\ref{fig:diag-cumulant}\ for the full DFT, and leave the
generalization to the other diagrams for the motivated reader.

\subsection{RG equations for the ``sub-model''}

The analomalous dimension $\zeta$ of the field $\vec{\phi}$ and dynamical
exponent read
\[
\zeta =-1-\frac{G^2}{8\pi} \quad \quad ;\quad \quad 
z=2+\frac{G^{2}}{8\pi }.
\]
 The scaling equations then follow : 
\begin{subequations}\label{RG-submodel}
\begin{align}
\partial_{l}G &=\frac{3}{16\pi }G^{3},\\
\partial_{l}U &=
-\frac{3}{8\pi }G^{2} U-\frac{1}{2\pi } U^{2}
-\frac{5}{2\pi }W^{2}-\frac{1}{8\pi }G^{4},\\ 
\partial_{l}V&=
-\frac{3}{8\pi }G^{2}V
-\frac{3}{2\pi }V^{2}-\frac{1}{\pi }W^{2}
-\frac{1}{\pi}UV
-\frac{1}{8\pi }G^{4},\\
\partial_{l}W&=
-\frac{3}{8\pi }G^{2}W-\frac{3}{2\pi
}(U+V)W+\frac{1}{4\pi }G^{4} .
\end{align}
\end{subequations}
 Note that these equations are a special case of the equations for the
full DFT presenetd below, with the only non-zero couplings
$G=-g_{0},U=u_{1},V=v_{1},W=w_{1}$.

\subsection{Tensorial notations for the full DFT}

To closely follow the study of the sub-model and the corresponding
1-loop expansion of Fig.~\ref{fig:diag-cumulant}, we need to translate
the definition of the full DFT into the tensorial coupling $G,U,V,W$,
corresponding to the 4 couplings of the sub-model.

With this idea in mind, the quadratic part, Eq.~(\ref{s2}), of the action
can be written as
\begin{eqnarray}
&&S^{(2)}=
-\int_{{\bf q},\omega} 
\vec{\phi}_{i}^{*}(-{\bf q},\omega) 
C({\bf q},\omega)
\vec{\phi}_{j}({\bf q},\omega),
\label{S2-tensor}\\
\nonumber 
&&\qquad C({\bf q},\omega) =\left[ C_{0}({\bf q},\omega)\delta_{ij}
+D_{ij}({\bf q})\right],
\end{eqnarray}
where $i=1,2$ or $x,y$, 
$C_{0}({\mathbf q},\omega)=(cq^{2}+iZ\omega)$, 
$c=c_{1}+c_{2},\delta c=c_{1}-c_{2}$ 
 and 
\begin{equation}
D_{ij}({\bf q})=\left(
\begin{array}{cc}
\delta c~
(q_{x}^{2}-q_{y}^{2})
&
d~ q_{x}q_{y } \\
d~ q_{x}q_{y } 
& 
-\delta c~
(q_{x}^{2}-q_{y}^{2})
\end{array}
 \right).
\end{equation}
Similarly, the cubic term 
\begin{eqnarray}
S^{(3)}_{int} & = &
\frac{i}{2} ~\int_{{\bf x}\tau}
\bigg\{  g_{0}
\vec{\phi}_{1}^{*}.\left(
\partial_{x} \vec{\phi}_{1}\times \vec{\phi}_{1}
- \vec{\phi}_{1}\times \partial_{x}\vec{\phi}_{1}\right)
\nonumber \\ & & + g_{1}
\vec{\phi}_{2}^{*}.\left(
\partial_{x} \vec{\phi}_{1}\times \vec{\phi}_{2}
- \vec{\phi}_{2}\times \partial_{x}\vec{\phi}_{1}\right)
\nonumber \\
& &+ g_{2}
\vec{\phi}_{1}^{*}.\left(
\partial_{x} \vec{\phi}_{2}\times \vec{\phi}_{2}
-\vec{\phi}_{2}\times \partial_{x} \vec{\phi}_{2}\right)
\nonumber \\ & & + g_{4}
\vec{\phi}_{2}^{*}.\left(
\partial_{x} \vec{\phi}_{2}\times \vec{\phi}_{1}
- \vec{\phi}_{1}\times \partial_{x}\vec{\phi}_{2}\right)
\bigg\}
\end{eqnarray}
can be rewritten as 
\begin{eqnarray}
S_{int}^{(3)} & = &
-\frac{1}{2}
\int_{{\bf q}_{1},{\bf q}_{2}}
\epsilon_{\mu \nu \rho }G_{ijk}({\bf q}_{1},{\bf q}_{2}) \nonumber \\
& & \left[
\vec{\phi}^{*}_{i,\mu }({\bf q}_{1}+{\bf q}_{2})
\vec{\phi}_{j,\nu }({\bf q}_{1})
\vec{\phi}_{k,\rho }({\bf q}_{2})
+ {\rm c.c.}
\right],
\label{S3-tensor}
\end{eqnarray}
where the coupling constant is
given by 
\begin{eqnarray}
&&G_{ijk}({\bf q}_{1},{\bf q}_{2})  = \\
\nonumber 
&&\qquad ~g_{0}~\delta_{111}(q_{1x}-q_{2x})
+g_{1}(\delta_{212}q_{1x}-\delta_{221}q_{2x}) \nonumber \\
&&\qquad  +g_{2}~\delta_{122}(q_{1x}-q_{2x})
+g_{4}(\delta_{221}q_{1x}-\delta_{212}q_{2x}) \nonumber \\
&&\qquad  +\left(x\rightarrow y,1\leftrightarrow 2 \right) \nonumber \\
&& \quad =  ~ 
\delta_{111}\left[g_{0}~(q_{1x}-q_{2x}) \right]
+\delta_{122}\left[g_{2}~(q_{1x}-q_{2x})\right] \nonumber \\
& & \qquad 
+\delta_{212}\left[g_{1}q_{1,x}-g_{4}q_{2,x}\right]
+\delta_{221}\left[g_{4}q_{1,x}-g_{1}q_{2,x}\right].
\nonumber 
\end{eqnarray}
Here we have used the notation 
$\delta_{111}=\delta_{i,1}\delta_{j,1}\delta_{k,1}$, etc.. 

With similar notations, the quartic part (\ref{s4}) reads 
\begin{equation}
S^{(4)}=\int_{{\bf x},\tau}
\vec{\phi}_{i}^{*}\vec{\phi}_{j}^{*}(U^{ij;kl}+V^{ij;kl}+W^{ij;k,l})
\vec{\phi}_{k}\vec{\phi}_{l},
\end{equation}
with the coupling tensors 
\begin{subequations}
\label{S4-tensor}
\begin{align}
U^{ij;kl}&=
 u_{1}(\delta^{ijkl}_{1111}+\delta^{ijkl}_{2222})
+u_{2}(\delta^{ijkl}_{2211}+\delta^{ijkl}_{1122})\nonumber \\
&
+u_{3}(\delta^{ijkl}_{2121}+\delta^{ijkl}_{1212})
+u_{4}(\delta^{ijkl}_{2112}+\delta^{ijkl}_{1221}),
\\
V^{ij;kl}&=
 v_{1}(\delta^{ijkl}_{1111}+\delta^{ijkl}_{2222})
+v_{2}(\delta^{ijkl}_{2211}+\delta^{ijkl}_{1122})\nonumber \\
&
+v_{3}(\delta^{ijkl}_{2121}+\delta^{ijkl}_{1221}
      +\delta^{ijkl}_{1212}+\delta^{ijkl}_{2112}),
\\
W^{ij;k,l}&=
 w_{1}(\delta^{ijkl}_{1111}+\delta^{ijkl}_{2222})
+w_{2}(\delta^{ijkl}_{2211}+\delta^{ijkl}_{1122})\nonumber \\
&
+w_{3}(\delta^{ijkl}_{2121}+\delta^{ijkl}_{1221}
      +\delta^{ijkl}_{1212}+\delta^{ijkl}_{2112}).
\end{align}
\end{subequations}

\subsection{Cubic diagram contribution}

 We choose to illustrate our method, and in particular our choice of
regularization, on the contribution of $G$ to order $G^{3}$ (see
figure below). As the cubic term is a specificity of the DFT we
developped in this paper, such a term is representative of the
differences between the renormalization of this field theory with
respect to conventional ones. The technical subtilities of the
regularization will also appear clearer on this diagram than on any
others, due to the explicit momentum dependance of the cubic term. 

\begin{figure}
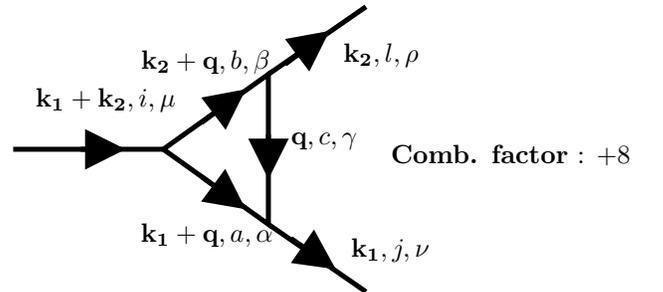

\centerline{
\psfrag{k1}{${\bf k_{2}},l,\rho$}
\psfrag{k2}{${\bf k_{1}},j,\nu$}
\psfrag{k3}{\hspace{-.6cm}${\bf k_{1}+q},a,\alpha$}
\psfrag{k4}{\hspace{-.6cm}${\bf k_{2}+q},b,\beta$}
\psfrag{k5}{${\bf q},c,\gamma$}
\psfrag{k6}{\hspace{-.5 cm}${\bf k_{1}+k_{2}},i,\mu$}
\diag{8mm}{5cm}{GGG-1}
\raisebox{1cm}{{\bf Comb. factor} : $+8$}}
\caption{Diagram corresponding to the renormalization of $G$ to order
$G^{3}$.}
\label{fig:G3}
\end{figure}
 
 With the momentum convention of Fig.~\ref{fig:G3}, we can write the
contribution of this $G^{3}$ diagram with the tensorial notations of
the previous part as  
\begin{align}\label{G3-contrib}
&\left(\frac{1}{2} \right)^{3}
\epsilon_{\mu \alpha \beta }\epsilon_{\beta \gamma \rho }
\epsilon_{\forall \gamma \alpha }
\\\nonumber 
&\quad 
\int_{{\bf q},\omega}C^{-1}({\bf k}_{1}-{\bf q},\Omega_{1}-\omega)
                     C^{-1}({\bf k}_{2}+{\bf q},\Omega_{2}+\omega)
                     C^{-1}({\bf q},\omega)\\\nonumber 
&\qquad \qquad 
G_{iab}({\bf k}_{1}-{\bf q},{\bf k}_{2}+{\bf q})
G_{bcl}({\bf q},{\bf k}_{2})
G_{jca}({\bf q},{\bf k}_{1}-{\bf q}). 
\end{align}
 The antisymmetric tensors $\epsilon_{\alpha \beta \gamma }$ come from
the vector product in the definition of the cubic term (\ref{s3}), 
 and the factor $\frac{1}{2}$ is part of the convention we chose for
the definition of the tensor $G$. In the above ${\bf q}$ integral, the
regularizaton  (\ref{cut-off}) is assumed for each of the three
$C^{-1}$ propagators.  
 By explicitly integrating over $\omega$ we can get rid of one of the
corresponding $t$ integral (see (\ref{cut-off})) : 
\begin{align*}
&\int_{\omega } C^{-1}({\bf k}_{1}-{\bf q},\Omega_{1}-\omega)
                     C^{-1}({\bf k}_{2}+{\bf q},\Omega_{2}+\omega)
                     C^{-1}({\bf q},\omega)\\
&=
\int_{\Lambda^{-2}}^{+\infty}\!\! dt_{1}
\int_{\Lambda^{-2}}^{+\infty}\!\! dt_{2}~ 
\exp \biggl[ -2(t_{1}+t_{2})
\left[{\bf q}+{\bf Q} \right]^{2}\\
&
-t_{1}(2{\bf Q}^{2}+{\bf k}_{1}^{2}+{\bf k}_{2}^{2}+i(\Omega_{1}+\Omega_{2}))
-t_{2}(2{\bf Q}^{2}+{\bf k}_{1}^{2}+i\Omega_{1})   
\biggr], 
\end{align*}
 with ${\bf Q}=\frac{1}{2}(\frac{t_{1}}{t_{1}+t_{2}}{\bf k}_{2}-{\bf k}_{1})$. 

To lighten the writing, let us use the notations 
$X=2Q^{2}+k_{1}^{2}+k_{2}^{2}+i(\Omega_{1}+\Omega_{2})$ 
and 
$Y=2Q^{2}+k_{1}^{2}+i\Omega_{1}$. These appear as factors 
respectively of $t_{1}$ and $t_{2}$ in the exponential in the result
of the $\omega$ integration. 
 After a change of variables, ${\bf q}\to {\bf q}+{\bf Q}$, the
remaining ${\bf q}$ dependant exponential becomes isotropic in {\bf q}. 
 Hence, upon expanding the $G^{3}$ product in (\ref{G3-contrib}), only
two terms do not vanish by ${\bf q}$-reflection symmetry : the 
$q_{x}^{2},q_{y}^{2}$ factor, and the constant factor. By power
counting, it is easily realized that the former 
is the only one proportional to $\ln \Lambda$ in the limit $\Lambda\to
\infty$.

 Due to the  change of variable, the $q_{x}^{2},q_{y}^{2}$ term appear
with two different $t$'s prefactors. 
 The first prefactor corresponds to the integral   
\begin{multline}\label{G3integral}
I_{1} = \int_{u}^{+\infty}\!\!\!\!dt_{1}\int_{u}^{+\infty}\!\!\!\!dt_{2}~
\frac{t_{1}}{t_{1}+t_{2}}
\\
\int_{{\bf q}} q_{x}^{2}
e^{-2(t_{1}+t_{2})q^{2}}\left( e^{-t_{1}X}e^{-t_{2}Y}\right). 
\end{multline}
 The ${\bf q}$ integral can then be performed explicitly, and we
obtain the result  
\[
I_{1} =\frac{1}{32\pi }
\int_{u}^{+\infty}dt_{1}\int_{u}^{+\infty}dt_{2}~
\frac{t_{1}}{(t_{1}+t_{2})^{3}}
\left(e^{-t_{1}X}e^{-t_{2}Y}\right). 
\]
 We are interested in the leading term in the limit of large cut-off
$\Lambda$ which corresponds to 
\begin{align*}
\partial_{u}I_{1}&=\frac{1}{32\pi }
\frac{1}{u}\int_{1}^{\infty}dx \frac{1}{(1+x)^{2}}
\Rightarrow 
I_{1}\simeq 
-\frac{1}{32\pi }\ln \Lambda . 
\end{align*}
 This defines the contribution from this part of the $G^{3}$
contraction.

The other integral corresponding to a constant prefactor of
$q_{x}^{2},q_{y}^{2}$ is 
\begin{align*}
I_{2}&= \int_{u}^{+\infty}\!\!\!dt_{1}\int_{u}^{+\infty}\!\!\!dt_{2}~
\int_{q} q_{x}^{2}
e^{-2(t_{1}+t_{2})q^{2}}e^{-t_{1}X}e^{-t_{2}Y}\\
&=-\frac{1}{16\pi }\ln \Lambda. 
\end{align*}

 We now have to collect all these results to explicitly contract the
three tensors  
\begin{align*}
&G_{iab}(k_{1}-q+Q,k_{2}+q-Q)
G_{bcl}(q-Q,k_{2})\\
&
G_{jca}(q-Q,k_{1}-q+Q), 
\end{align*}
 and simplify them with the results of the two above integrals. 
 This is most conveniently done using a numerical program like {\it
e.g} Mathematica.  
 The resulting $G^{3}$ contribution to the $G$'s is embedded in the
results of the next subsection.

\subsection{Explicit RG equations}

 Armed with this example, one can simply repeat this procedure for
each of the diagram of figure \ref{fig:diag-cumulant}. 
 The final explicit result reads, in terms of the original
coefficients, the following equations. Although the
 only useful form of them may be a numerical routine, we include them
for completeness and for the brave and bold readers. 

\startlargeeq

\begin{eqnarray*}
\partial_{l}\delta c &=&
- \frac{1}{8\pi }\left(g_{1}-g_{4} \right)^{2}
+ \frac{1}{8\pi }
\left(
-\frac{1}{2}\delta c (3g_{0}^2+g_{2}^2-g_1^2+g_4^2+4g_{1}g_{4})
+\frac{d}{2}(g_1-g_4)(g_{0}-g_{2})\right), \\
\partial_{l}d&=&
- \frac{1}{8\pi }\left(g_{1}-g_{4} \right)^{2}
- \frac{1}{8\pi }
\left(\frac{1}{2}\delta c(g_{4}^{2}-g_{1}^{2})
+d(g_{0}^{2}+g_{2}^{2}+2g_{1}g_{4}+g_{0}g_{4}+g_{1}g_{2}) \right), 
\end{eqnarray*}

\begin{eqnarray*}
\partial_{l}g_{0}&=&
\left( -\frac{1}{4\pi }\left(g_{0}^{2}+g_{2}^{2} \right)
-\frac{1}{16\pi}\left(g_{1}+g_{4}\right)^{2}
\right)g_{0}
+\frac{1}{32\pi }\biggl[
14 g_{0}^3 + \frac{5}{2}g_{0} g_1^2 + 2g_1^3 + g_{0} g_1g_{2} +
    g_1^2g_{2} + 7g_1g_{2}^2 + 6g_{0} g_1g_4 \\
&&\qquad \qquad \qquad \qquad \qquad \qquad \qquad \qquad \qquad \qquad
+ \frac{11}{2} g_1^2g_4 +
    g_{0} g_{2}g_4 + g_1g_{2}g_4 +5g_{2}^2g_4 + \frac{7}{2}g_{0} g_4^2 +
    5g_1g_4^2 + \frac{3}{2}g_4^3
\biggr],\\
\partial_{l}g_{1}&=&
\left( -\frac{1}{4\pi }\left(g_{0}^{2}+g_{2}^{2} \right)
-\frac{1}{16\pi}\left(g_{1}+g_{4}\right)^{2}
\right)g_{1}
-\frac{1}{4\pi } (g_{1}-g_{4})(u_{2}-u_{4})
+\frac{1}{2\pi } (g_{1}-g_{4})(w_{2}-w_{3})\\
&&+\frac{1}{32\pi }\biggl[
8g_{0}^2g_1 - \frac{1}{2}g_{0} g_1^2 + 2g_1^3 + g_{0} g_1g_{2} +
    g_1^2g_{2} + 10g_{0} g_{2}^2 + 7g_1g_{2}^2 
+ 2g_{0}^2g_4 +
    12g_{0} g_1g_4 \\
&&\qquad \qquad + \frac{13}{2}g_1^2g_4 + g_{0} g_{2}g_4
    + g_1g_{2}g_4 -
    g_{2}^2g_4
 + \frac{9}{2}g_{0} g_4^2 + 3g_1g_4^2
    - \frac{3}{2}g_4^3
\biggr],\\
\partial_{l}g_{2}&=&
\left( -\frac{1}{4\pi }\left(g_{0}^{2}+g_{2}^{2} \right)
-\frac{1}{16\pi}\left(g_{1}+g_{4}\right)^{2}
\right)g_{2}
+\frac{1}{32\pi }\biggl[
-\frac{1}{2}g_{0} g_1^2 - 2g_{0}^2g_{2} + 12g_{0} g_1g_{2} +
    8g_1^2g_{2} + g_{0} g_1g_4 + \frac{3}{2}g_1^2g_4 \\
&&\qquad \qquad\qquad \qquad\qquad \qquad\qquad \qquad\qquad \qquad
    + 14g_{0} g_{2}g_4 +
    12g_1g_{2}g_4 + \frac{3}{2}g_{0} g_4^2 + g_1g_4^2 + 8g_{2}g_4^2
   -\frac{1}{2}g_4^3
\biggr],\\
\partial_{l}g_{4}&=&
\left( -\frac{1}{4\pi }\left(g_{0}^{2}+g_{2}^{2} \right)
-\frac{1}{16\pi}\left(g_{1}+g_{4}\right)^{2}
\right)g_{4}
+\frac{1}{4\pi } (g_{1}-g_{4})(u_{2}-u_{4})
-\frac{1}{2\pi } (g_{1}-g_{4})(w_{2}-w_{3})\\
&&+\frac{1}{32\pi }\biggl[
\frac{19}{2}g_{0} g_1^2 + 2g_{0}^2g_{2} + 2g_{0} g_1g_{2} - 
    2g_1^2g_{2} + 4g_{0} g_{2}^2 + 6g_1g_{2}^2 \\
&&\qquad \qquad + 4g_{0}^2g_4 +
    7g_{0} g_1g_4 - \frac{5}{2}g_1^2g_4 + 2g_1g_{2}g_4 
    + 14g_{2}^2g_4 
+  \frac{11}{2}g_{0} g_4^2 + g_1g_4^2 + \frac{7}{2}g_4^3
\biggr],
\end{eqnarray*}

\begin{eqnarray*}
\partial_{l}u_{1}&=&
\left(-\frac{3}{8\pi }\left(g_{0}^{2}+g_{2}^{2} \right)
-\frac{1}{8\pi}\left(g_{1}+g_{4}\right)^{2}
 \right)u_{1}
-\frac{1}{2\pi }(u_{1}^2 + u_{3}^2)
-\frac{3}{2\pi }(w_{1}^2 + w_{2}^2)
-\frac{1}{\pi }(w_{1}^2 + w_{3}^2)
\\
&&
-\frac{1}{64\pi }
\biggl(
12g_{0}^4 + 4g_{0}^2g_{1}^2 + 3g_{1}^4 + 4g_{0}^2g_{1}g_{2} + 
    8g_{0}g_{1}g_{2}^2 + 3g_{1}^2g_{2}^2 + 4g_{0}^2g_{1}g_{4} + 
    2g_{0}g_{1}^2g_{4} + 6g_{1}^3g_{4}\\
&&\qquad \qquad 
 + 4g_{0}^2g_{2}g_{4} +
4g_{1}^2g_{2}g_{4} +  
    10g_{1}g_{2}^2g_{4} + 4g_{0}g_{1}g_{4}^2 + 3g_{1}^2g_{4}^2 + 
    4g_{1}g_{2}g_{4}^2 + 19g_{2}^2g_{4}^2 + 2g_{0}g_{4}^3
\biggr),
\\
\partial_{l}u_{2}&=&
\left(-\frac{3}{8\pi }\left(g_{0}^{2}+g_{2}^{2} \right)
-\frac{1}{8\pi}\left(g_{1}+g_{4}\right)^{2}
 \right)u_{2}
-\frac{1}{2\pi }(u_{2}^2 + u_{4}^2)
-\frac{3}{2\pi }(2 w_{1} w_{2})
-\frac{1}{\pi }(w_{2}^2 + w_{3}^2)\\
&&
-\frac{1}{64\pi }
\biggl(
17g_{0}^2g_{1}^2 + 2g_{1}^3g_{2} + 4g_{0}g_{1}g_{2}^2 + 4g_{2}^4 + 
    14g_{0}^2g_{1}g_{4} + 4g_{0}g_{1}^2g_{4} + 8g_{0}^2g_{2}g_{4} + 
    4g_{1}^2g_{2}g_{4} + 4g_{0}g_{2}^2g_{4}\\
&&\qquad \qquad 
 + 12g_{1}g_{2}^2g_{4} + g_{0}^2g_{4}^2 + 
    4g_{0}g_{1}g_{4}^2 + g_{1}^2g_{4}^2 + 2g_{1}g_{2}g_{4}^2 + 
12g_{2}^2g_{4}^2 + 
    2g_{1}g_{4}^3 + g_{4}^4
\biggr),
\\
\partial_{l}u_{3}&=&
\left(-\frac{3}{8\pi }\left(g_{0}^{2}+g_{2}^{2} \right)
-\frac{1}{8\pi}\left(g_{1}+g_{4}\right)^{2}
 \right)u_{3}
-\frac{1}{2\pi }(2 u_{1} u_{3})
-\frac{3}{2\pi }(2w_{3}^2)
-\frac{1}{\pi }(2 w_{2} w_{3})\\
&&
-\frac{1}{64\pi }
\biggl(
g_{1}^4 + 4g_{0}g_{1}^2g_{2} + 4g_{0}^2g_{2}^2 + 28g_{0}g_{1}g_{2}^2 + 
    g_{1}^2g_{2}^2 + 10g_{0}g_{1}^2g_{4} + 2g_{1}^3g_{4} +
8g_{0}g_{1}g_{2}g_{4} + 
    4g_{0}g_{2}^2g_{4}\\
&&\qquad \qquad 
 + 2g_{1}g_{2}^2g_{4} + 4g_{0}^2g_{4}^2 + 
    16g_{0}g_{1}g_{4}^2 + g_{1}^2g_{4}^2 + 4g_{0}g_{2}g_{4}^2 + g_{2}^2g_{4}^2 
+ 
    6g_{0}g_{4}^3
\biggr),
\\
\partial_{l}u_{4}&=&
\left(-\frac{3}{8\pi }\left(g_{0}^{2}+g_{2}^{2} \right)
-\frac{1}{8\pi}\left(g_{1}+g_{4}\right)^{2}
 \right)u_{4}
-\frac{1}{2\pi }(2 u_{2} u_{4})
-\frac{3}{2\pi }(2w_{3}^2)
-\frac{1}{\pi }(2 w_{1} w_{3})\\
&&
-\frac{1}{64\pi }
\biggl(
g_{0}^2g_{1}^2 + 12g_{0}^2g_{1}g_{2} + 4g_{0}g_{1}^2g_{2} + 2g_{1}^3g_{2} + 
    4g_{0}^2g_{2}^2 + 4g_{1}^2g_{2}^2 + 2g_{0}^2g_{1}g_{4} + 
    20g_{0}^2g_{2}g_{4} \\
&&\qquad \qquad 
+ 8g_{0}g_{1}g_{2}g_{4} + 16g_{1}^2g_{2}g_{4} + 
    g_{0}^2g_{4}^2 + g_{1}^2g_{4}^2 + 4g_{0}g_{2}g_{4}^2 + 14g_{1}g_{2}g_{4}^2 
+ 
    2g_{1}g_{4}^3 + g_{4}^4
\biggr),
\end{eqnarray*}

\begin{eqnarray*}
\partial_{l}v_{1}&=&
\left(-\frac{3}{8\pi }\left(g_{0}^{2}+g_{2}^{2} \right)
-\frac{1}{8\pi}\left(g_{1}+g_{4}\right)^{2}
 \right)v_{1}
-\frac{1}{\pi }(w_{1}^2 + w_{3}^2)
-\frac{1}{\pi }(u_{1} v_{1}+u_{3} v_{2})
-\frac{3}{2\pi }(v_{1}^{2}+v_{2}^{2})
\\
&&
-\frac{1}{64\pi }
\biggl(
12g_{0}^4 + 4g_{0}^2g_{1}^2 + 3g_{1}^4 + 4g_{0}^2g_{1}g_{2} + 
    8g_{0}g_{1}g_{2}^2 + 3g_{1}^2g_{2}^2 + 4g_{0}^2g_{1}g_{4} + 
    2g_{0}g_{1}^2g_{4} + 6g_{1}^3g_{4} + 4g_{0}^2g_{2}g_{4} \\
&&\qquad \qquad 
  + 4g_{1}^2g_{2}g_{4} + 
    10g_{1}g_{2}^2g_{4} + 4g_{0}g_{1}g_{4}^2 + 3g_{1}^2g_{4}^2 + 
    4g_{1}g_{2}g_{4}^2 + 19g_{2}^2g_{4}^2 + 2g_{0}g_{4}^3
\biggr),
\\
\partial_{l}v_{2}&=&
\left(-\frac{3}{8\pi }\left(g_{0}^{2}+g_{2}^{2} \right)
-\frac{1}{8\pi}\left(g_{1}+g_{4}\right)^{2}
 \right)v_{2}
-\frac{1}{\pi }(w_{2}^2 + w_{3}^2)
-\frac{1}{\pi }(u_{3} v_{1}+u_{1} v_{2})
-\frac{3}{2\pi }(2v_{1}v_{2})\\
&&
-\frac{1}{64\pi }
\biggl(
g_{0}^2g_{1}^2 + 12g_{0}^2g_{1}g_{2} + 4g_{0}g_{1}^2g_{2} + 2g_{1}^3g_{2} + 
    4g_{0}^2g_{2}^2 + 4g_{1}^2g_{2}^2 + 2g_{0}^2g_{1}g_{4} +
    20g_{0}^2g_{2}g_{4} + 8g_{0}g_{1}g_{2}g_{4}\\
&&\qquad \qquad  + 16g_{1}^2g_{2}g_{4} + 
    g_{0}^2g_{4}^2 + g_{1}^2g_{4}^2 + 4g_{0}g_{2}g_{4}^2 + 14g_{1}g_{2}g_{4}^2 
+ 
    2g_{1}g_{4}^3 + g_{4}^4
\biggr),
\\
\partial_{l}v_{3}&=&
\left(-\frac{3}{8\pi }\left(g_{0}^{2}+g_{2}^{2} \right)
-\frac{1}{8\pi}\left(g_{1}+g_{4}\right)^{2}
 \right)v_{3}
-\frac{1}{\pi }(w_{1} w_{3}+w_{2} w_{3})
-\frac{1}{\pi }(u_{2} v_{3}+u_{4} v_{3})
-\frac{3}{2\pi }2v_{3}^{2}\\
&&
-\frac{1}{64\pi }
\biggl(
\frac{17}{2}g_{0}^2g_{1}^2 + \frac{1}{2}g_{1}^4 + 2g_{0}g_{1}^2g_{2} +
g_{1}^3g_{2} + 
    2g_{0}^2g_{2}^2 + 16g_{0}g_{1}g_{2}^2 + \frac{1}{2}g_{1}^2g_{2}^2 + 
2g_{2}^4 + 
    7g_{0}^2g_{1}g_{4} \\
&&\qquad \qquad 
+ 7g_{0}g_{1}^2g_{4} + g_{1}^3g_{4} + 4g_{0}^2g_{2}g_{4} + 
    4g_{0}g_{1}g_{2}g_{4} + 2g_{1}^2g_{2}g_{4} + 4g_{0}g_{2}^2g_{4} + 
    7g_{1}g_{2}^2g_{4} \\
&&\qquad \qquad 
+ \frac{5}{2}g_{0}^2g_{4}^2 + 10g_{0}g_{1}g_{4}^2 + 
    g_{1}^2g_{4}^2 + 2g_{0}g_{2}g_{4}^2 + g_{1}g_{2}g_{4}^2 + 
    \frac{13}{2} g_{2}^2g_{4}^2 + 3g_{0}g_{4}^3 + g_{1}g_{4}^3 + \frac{1}{2} 
g_{4}^4 
\biggr),
\end{eqnarray*}

\begin{eqnarray*}
\partial_{l}w_{1}&=&
\left(-\frac{3}{8\pi }\left(g_{0}^{2}+g_{2}^{2} \right)
-\frac{1}{8\pi}\left(g_{1}+g_{4}\right)^{2}
 \right)w_{1}\\
&&
-\frac{1}{64\pi }
\biggl[
12g_{0}^4 + 4g_{0}^2g_{1}^2 + 4g_{0}^2g_{1}g_{2} + 4g_{0}g_{1}^2g_{2} + 
    6g_{1}^3g_{2} + 4g_{0}g_{2}^3
 + 4g_{0}^2g_{1}g_{4} +2g_{0}g_{1}^2g_{4}
 + 
    4g_{0}^2g_{2}g_{4} + 16g_{1}^2g_{2}g_{4} + 4g_{1}g_{2}^2g_{4} \\
&&\qquad \qquad  
+  4g_{0}g_{1}g_{4}^2 + 10g_{1}g_{2}g_{4}^2 + 16g_{2}^2g_{4}^2 + 2g_{0}g_{4}^3
\biggr]
\\
&&-\frac{1}{\pi }(u_{1} w_{1}+u_{3} w_{3})
-\frac{1}{2\pi }(u_{1} w_{1}+u_{3} w_{2})
-\frac{3}{2\pi }(v_{1}w_{1}+v_{2}w_{2})
-\frac{1}{\pi }(v_{1} w_{1}+v_{2} w_{3}),
\\
\partial_{l}w_{2}&=&
\left(-\frac{3}{8\pi }\left(g_{0}^{2}+g_{2}^{2} \right)
-\frac{1}{8\pi}\left(g_{1}+g_{4}\right)^{2}
 \right)w_{2}\\
&&
-\frac{1}{64\pi }
\biggl[
7g_{0}^2g_{1}^2 + g_{1}^4 + 10g_{0}^2g_{1}g_{2} + 2g_{0}g_{1}^2g_{2} +
    2g_{0}g_{1}g_{2}^2 + g_{1}^2g_{2}^2 
+ 4g_{1}g_{2}^3 + 8g_{0}^2g_{1}g_{4}
+ 2g_{0}g_{1}^2g_{4}
 + 2g_{1}^3g_{4} + 14g_{0}^2g_{2}g_{4} \\
&&
+4g_{0}g_{1}g_{2}g_{4} +  
    6g_{1}^2g_{2}g_{4} + 2g_{0}g_{2}^2g_{4}\ + 8g_{1}g_{2}^2g_{4}
+g_{0}^2g_{4}^2  
+ 2g_{0}g_{1}g_{4}^2 
+ 2g_{1}^2g_{4}^2 + 2g_{0}g_{2}g_{4}^2 + 
    6g_{1}g_{2}g_{4}^2 + 7g_{2}^2g_{4}^2 + 2g_{1}g_{4}^3 + g_{4}^4 
\biggr]
\\  
&&-\frac{1}{\pi }(u_{2} w_{2}+u_{4} w_{3})
-\frac{1}{2\pi }(u_{3} w_{1}+u_{1} w_{2})
-\frac{3}{2\pi }(v_{2}w_{1}+v_{1}w_{2})
-\frac{1}{\pi }(v_{2} w_{1}+v_{1} w_{3}),
\\
\partial_{l}w_{3}&=&
\left(-\frac{3}{8\pi }\left(g_{0}^{2}+g_{2}^{2} \right)
-\frac{1}{8\pi}\left(g_{1}+g_{4}\right)^{2}
 \right)w_{3}\\
&&
-\frac{1}{64\pi }
\biggl[
\frac{7}{2}g_{0}^2g_{1}^2 + g_{1}^4\/2 + 5g_{0}^2g_{1}g_{2} + 
    9g_{0}g_{1}^2g_{2} + g_{1}^3g_{2} + 2g_{0}^2g_{2}^2 
 + 3g_{0}g_{1}g_{2}^2 + 
    \frac{1}{2}g_{1}^2g_{2}^2 + 6g_{0}g_{2}^3 
+ 2g_{1}g_{2}^3 +
4g_{0}^2g_{1}g_{4} +  
    6g_{0}g_{1}^2g_{4} \\
&& \qquad + g_{1}^3g_{4} + 7g_{0}^2g_{2}g_{4} +
6g_{0}g_{1}g_{2}g_{4} +   
    5g_{1}^2g_{2}g_{4} + 3g_{0}g_{2}^2g_{4} 
+ 4g_{1}g_{2}^2g_{4} + 
    \frac{5}{2}g_{0}^2g_{4}^2 + 9g_{0}g_{1}g_{4}^2 + g_{1}^2g_{4}^2 + 
    3g_{0}g_{2}g_{4}^2 + 4g_{1}g_{2}g_{4}^2 \\
&&\qquad 
+ \frac{7}{2}g_{2}^2g_{4}^2+ 
    3g_{0}g_{4}^3 + g_{1}g_{4}^3 + \frac{1}{2}g_{4}^4
\biggr]\\
&&-\frac{1}{\pi }
\frac{1}{2}(u_{3} w_{1} + u_{4} w_{2} + u_{1} w_{3} + u_{2} w_{3})
-\frac{1}{2\pi }(u_{2} w_{3}+u_{4} w_{3})
-\frac{3}{2\pi }(2v_{3}w_{3})
-\frac{1}{\pi }(v_{3} w_{2}+v_{3} w_{3}). 
\end{eqnarray*}
\begin{multicols}{2}

\end{multicols}

\begin{references}
\bibitem{comment-Heinsenberg}
Note that the SS
dimer state is thus an exact eigenstate of the 2D Heisenberg
antiferromagnet (however, like the fully-polarized ferromagnetic
eigenstate, it is highly excited for $J_1=0$).
\bibitem{comment} We deviate slightly from Ref.~\onlinecite{Sachdev90}\ by not 
  introducing redundant ``singlet'' operators.  These were intended
  technically to maintain the hard-core constraint, thereby
  introducing an unphysical U(1) gauge symmetry.  For the general
  coarse-grained Landau theory considerations employed here, they are
  unnecessary.
\end{references}
\end{document}